%% file: main.tex
\documentclass[11pt,
  a4paper, oneside,
  parskip=half, %
  ngerman,
  english]{scrbook}

\usepackage[ngerman, english]{babel} %
\usepackage{titling}
\usepackage{geometry}
\usepackage{enumitem}
\usepackage[nottoc,numbib]{tocbibind}

\title{A Framework for Preserving Privacy and Cybersecurity in Brain-Computer Interfacing Applications}
\author{NeuroMentum AI GmbH}

\input{setup}

\begin{document}
    \pagestyle{empty} %
    \hypersetup{pageanchor=false}

\input{chapters/0_0-titlepage_en}
    \pagestyle{headings}
    \frontmatter  %
    \input{chapters/0_2-abstract}
    \input{chapters/0_3-acknowledgments.tex}
    \tableofcontents
    \listoffigures
    \glsaddall
    \printglossary[type=\acronymtype,title=Acronyms]
    \hypersetup{pageanchor=true}  %

    \mainmatter  %
    \input{chapters/1-introduction}

    \input{chapters/2-definition}
    \input{chapters/3-bci-concepts}
    \input{chapters/4-relatedwork}

    \input{chapters/5-architecture}

    \input{chapters/6-privacy-designs-strategies}

    \input{chapters/7-application}

    \input{chapters/8-conclusions}

    \bibliographystyle{ieeetr}
    \bibliography{bib/main}

\end{document}

%% file: setup.tex
    \usepackage[T1]{fontenc}  %
    \usepackage[utf8x]{inputenc}
    \usepackage{scrhack}

    \usepackage[labelfont=bf, labelsep=colon, format=hang, textfont=singlespacing]{caption}
    \usepackage{chngcntr}  %
    \counterwithout{equation}{chapter}
    \counterwithout{figure}{chapter}
    \counterwithout{table}{chapter}

    \setkomafont{chapter}{\normalfont\bfseries\huge}

    \usepackage{setspace}  %
    \usepackage[dvipsnames]{xcolor}  %
    \usepackage{array}  %
    \usepackage{graphicx}  %
    \usepackage{subfig}  %
    \usepackage{amsmath}  %
    \usepackage{amsthm}   %
    \usepackage{calc}  %
    \usepackage[unicode=true,bookmarks=true,bookmarksnumbered=true,
                bookmarksopen=true,bookmarksopenlevel=1,breaklinks=false,
                pdfborder={0 0 0},backref=false,colorlinks=false]{hyperref}
    \usepackage{etoolbox} %

    \usepackage{algorithm,algpseudocode}
    \usepackage{bm}  %
    \usepackage{tikz}
    \usetikzlibrary{positioning}  %

    \usepackage{ifthen}

    \usepackage{enumitem}
    \usepackage{booktabs}

    \usepackage{lipsum}
    
    \usepackage{nameref}
    \usepackage{float}
     \usepackage{dirtytalk}

    \usepackage[numbers,sort&compress]{natbib}
    
    \usepackage{multirow}
    \usepackage{tablefootnote}
    \usepackage{tikz,lipsum,lmodern}
    \usepackage[dvipsnames]{xcolor}
    \usepackage[most]{tcolorbox}

    \renewcommand{\eqref}[1]{Equation~(\ref{#1})}

    \definecolor{darkgreen}{rgb}{0.0, 0.5, 0.0}

    \definecolor{UniBlue}{RGB}{0, 74, 153}
    \definecolor{UniRed}{RGB}{193, 0, 42}
    \definecolor{UniGrey}{RGB}{154, 155, 156}

    \usepackage[acronym,nomain,nonumberlist]{glossaries}
    \makeglossaries
    
    \newacronym{eeg}{EEG}{Electroencephalography}
    \newacronym{bci}{BCI}{Brain-Computer Interface}
    \newacronym{vr}{VR}{Virtual Reality}
    \newacronym{sdk}{SDK}{Software Development Kit}
    \newacronym{api}{API}{Application Programming Interface}
    \newacronym{trl}{TRL}{Technology Readiness Level}
    \newacronym{gdpr}{GDPR}{General Data Protection Regulation}
    \newacronym{fipps}{FIPPs}{Fair Information Practice Principles}
    \newacronym{cns}{CNS}{Central Nervous System}
    \newacronym{meg}{MEG}{Magnetoencephalography}
    \newacronym{fnirs}{fNIRS}{functional Near-Infrared Spectroscopy}
    \newacronym{fmri}{fMRI}{functional Magnetic Resonance Imaging}
    \newacronym{ecog}{ECoG}{Electrocorticography}
    \newacronym{lfp}{LFP}{Local Field Potential}
    \newacronym{sua}{SUA}{Single Unit Activity}
    \newacronym{mua}{MUA}{Multi Unit Activity}
    \newacronym{ssvep}{SSVEP}{Steady State Visually Evoked Potential}
    \newacronym{ml}{ML}{Machine Learning}
    \newacronym{svm}{SVM}{Support Vector Machine}
    \newacronym{rlda}{RLDA}{Regularized Linear Discriminant Analysis}
    \newacronym{dl}{DL}{Deep Learning}
    \newacronym{asic}{ASIC}{Application-Specific Integrated Circuit}
    \newacronym{ncd}{NCD}{Near Control Device}
    \newacronym{wipo}{WIPO}{World International Property Organization}
    \newacronym{opm}{OPMs}{Optically Pumped Magnetometers}
    \newacronym{ann}{ANNs}{Artificial Neural Networks}
    \newacronym{ppml}{PPML}{Privacy-Preserving Machine Learning}
    \newacronym{ddos}{DDoS}{Distributed Denial-of-Service}
    \newacronym{erp}{ERP}{Event-Related Potential}
    \newacronym{poc}{PoC}{Proof-of-Concept}
    \newacronym{av}{AV}{Attack Vector}
    \newacronym{dfd}{DFD}{Data Flow Diagram}
    \newacronym{aws}{AWS}{Amazon Web Services}
    \newacronym{pet}{PETs}{Privacy Enhancing Technologies}
    \newacronym{owasp}{OWASP}{Open Web Application Security Project}
    \newacronym{oecd}{OECD}{Organisation for Economic Co-operation and Development}
    \newacronym{gan}{GAN}{Generative Adversarial Network}
    \newacronym{lfc}{LFC}{Low frequency component}
    \newacronym{smc}{SMC}{Secure Multiparty Computation}
    \newacronym{tte}{TEEs}{Trusted Execution Environments}
    \newacronym{ern}{ERN}{Error-Related Negativity}
    \newacronym{dsgvo}{DSGVO}{Datenschutzgrundverordnung}
    \newacronym{bdsg}{BDSG}{Bundesdatenschutzgesetz}
    \newacronym{ai}{AI}{Artificial Intelligence}
    \newacronym{hipaa}{HIPAA}{Health Insurance Portability and Accountability Act}

    \let\mySection\section\renewcommand{\section}{\suppressfloats[t]\mySection}
    \let\mySubSection\subsection\renewcommand{\subsection}{\suppressfloats[t]\mySubSection}

    \hypersetup{pdftitle={\thetitle},
                pdfauthor={\theauthor},
                pdfsubject={},
                pdfkeywords={},
                pdfpagelayout=OneColumn, pdfnewwindow=true, pdfstartview=XYZ, plainpages=false}

    \tikzset{>=latex}  %

    \algnewcommand\algorithmicforeach{\textbf{foreach}}
    \algdef{S}[FOR]{ForEach}[1]{\algorithmicforeach\ #1\ \algorithmicdo}

    \setlist{itemsep=.5em}

	\usepackage{ifthen} %
	\newcounter{todos}
	\newcounter{extends}
	\newcounter{drafts}

%% file: chapters/0_0-titlepage_en.tex
\begin{titlepage}
\begin{center}

\newcommand{\HorizontalLine}{\rule{\linewidth}{0.3mm}}

\HorizontalLine \\[0.4cm]
{ \huge \bfseries \thetitle }
\HorizontalLine \\[1.5cm]

\bigskip
\bigskip

\includegraphics[width=1.\textwidth]{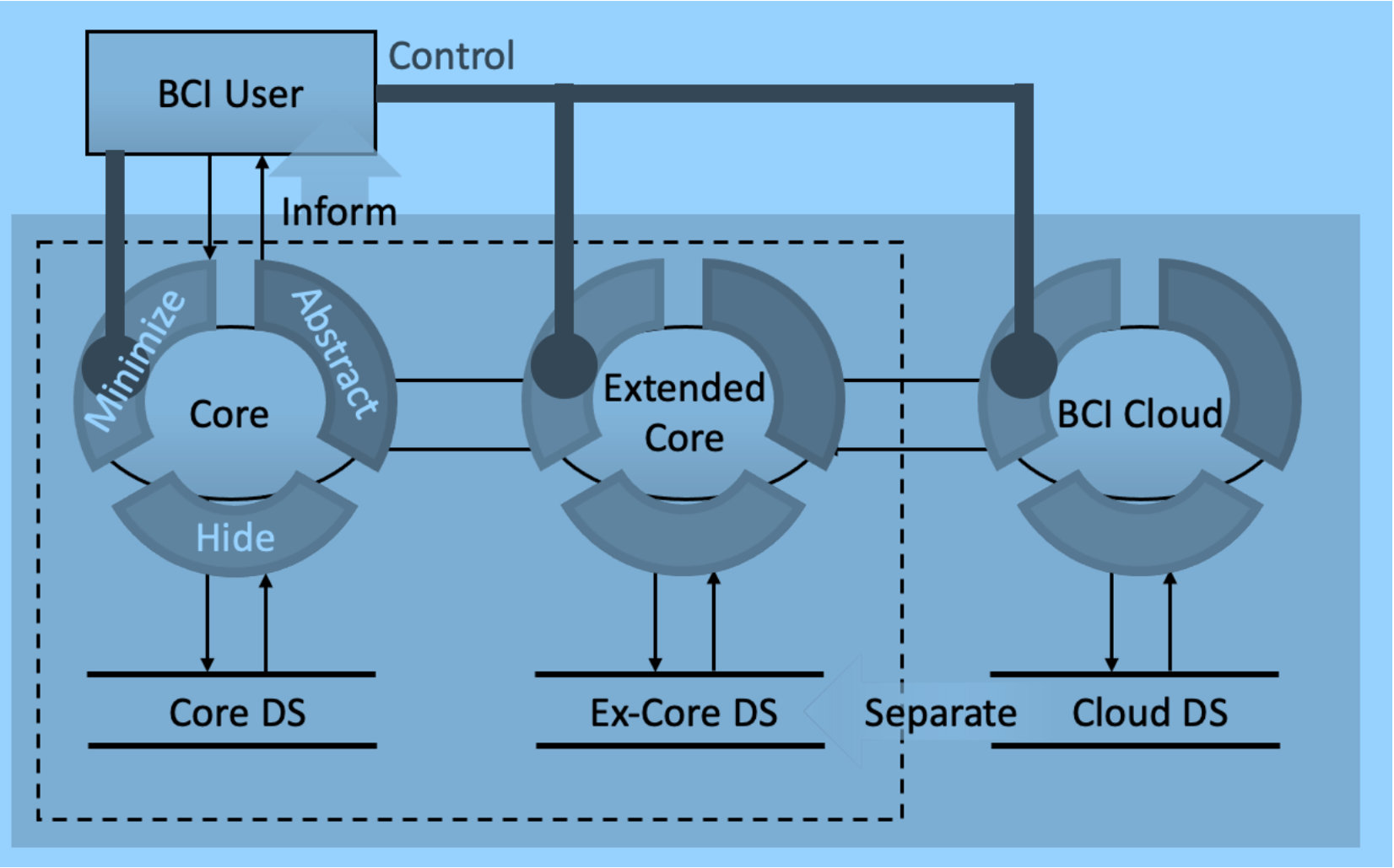}
\vfill  %

\Large {
    
    Maryna Kapitonova, NeuroMentum AI, Freiburg\\
    Philipp Kellmeyer, Human-Technology Interaction Lab, Freiburg\\
    Simon Vogt, Innovation for Cybersecurity, Halle\\
    Tonio Ball, NeuroMentum AI, Freiburg\\

    \bigskip

    Version 1.1 --- September 16\textsuperscript{th}, 2022\\
    \textcopyright~Agentur für Innovation in der Cybersicherheit GmbH (Innovation for Cybersecurity)\\
}
\end{center}
\end{titlepage}

%% file: chapters/0_2-abstract.tex
\chapter*{Abstract}
\acrlong{bci}s (\acrshort{bci}s) comprise a rapidly evolving field of technology with the potential of far-reaching impact in domains ranging from medical over industrial to artistic, gaming, and military. \acrshort{bci}s provide technical interfaces with recording and/or stimulation functionality to connect the brain with computer systems running \verb|"decoders"| for online analysis of the recorded brain signals. This online analysis in turn can inform various effectors such as robots, vehicles (brain-to-vehicle interfaces), brain stimulation devices, or computer games (neurogaming). Today, these emerging \acrshort{bci} applications are typically still at early technology readiness levels, but because \acrshort{bci}s create novel, technical communication channels for the human brain, they have raised privacy and security concerns. In particular, as brain data contain personal information, adversaries may utilize \acrshort{bci}s to compromise  brain privacy. There are first empirical proofs-of-principle that such attacks are indeed possible, possibly foreshadowing a next level of privacy and cybersecurity threats targeting the brain by neurotechnological means. To mitigate such risks, a large body of countermeasures has been proposed in the literature, but a general framework is lacking which would describe how privacy and security of \acrshort{bci} applications can be protected by design, i.e., already as an integral part of the early \acrshort{bci} design process, in a systematic manner, and allowing suitable depth of analysis for different contexts such as commercial \acrshort{bci} product development vs. academic research and lab prototypes. 

Here we propose the adoption of recent systems-engineering methodologies for privacy threat modeling, risk assessment, and privacy engineering to the \acrshort{bci} field. These methodologies address privacy and security concerns in a more systematic and holistic way than previous approaches, and provide reusable patterns on how to move from principles to actions. We apply these methodologies to \acrshort{bci} processes and data flows and derive a generic, extensible, and actionable framework for brain-privacy-preserving cybersecurity in \acrshort{bci} applications. This framework is designed for flexible application to the wide range of current and future \acrshort{bci} applications. We also propose a range of novel privacy-by-design features for \acrshort{bci}s, with an emphasis on features promoting \acrshort{bci} transparency as a prerequisite for informational self-determination of \acrshort{bci} users, as well as design features for ensuring \acrshort{bci} user autonomy. We anticipate that our framework will contribute to the development of privacy-respecting, trustworthy \acrshort{bci} technologies.

%% file: chapters/0_3-acknowledgments.tex
\chapter*{Acknowledgments}

The primary development of this document was funded by the Agentur für Innovation in der Cybersicherheit GmbH (Innovation for Cybersecurity).

We thank Paula Vieweg for her helpful comments on the manuscript.

\chapter*{Disclaimer}

As part of the project \emph{"Secure Neural Human-Machine Interaction"} of the German Agentur für Innovation in der Cybersicherheit GmbH (Innovation for Cybersecurity), this document outlines a formalism to design neural communication between human and machine in such a way that fundamental personal rights, data security and data integrity are ensured in the acquisition, analysis and interpretation of neural data.

The views and conclusions contained herein are those of the authors and should not be interpreted as necessarily representing the official policies or endorsements, either expressed or implied, of the German Agentur für Innovation in der Cybersicherheit GmbH (Innovation for Cybersecurity), or the German Government.

%% file: chapters/1-introduction.tex
\chapter{Introduction}\label{chap:introduction}

\acrlong{bci}s (\acrshort{bci}s) are an emerging class of technology enabling direct uni- or bidirectional communication between brains and technical devices~\cite{rao_brain-computer_2013, wolpaw_braincomputer_2012}. \acrshort{bci}s can measure and convert brain activity to outputs for control of technical effectors, or to perform brain stimulation. \acrshort{bci} applications have been proposed for a wide range of domains from medical and well-being, work and employment, productivity, cognitive  enhancement, education, artistic, neurogaming and entertainment including novel interfaces to \acrfull{vr}, neuromarketing, \acrshort{bci}-informed smart homes and smart cities, to security and military-related \acrshort{bci} applications. 

This broad spectrum is the result of a rapid growth of the \acrshort{bci} field: After decades of research confined to a (relatively) small academic group, today a large research community has established itself around \acrshort{bci}-related topics, including the legal, ethical, and societal aspects of \acrshort{bci} technologies. Also the commercialization efforts for medical and consumer \acrshort{bci} systems are intensifying. The global market for \acrshort{bci}s in medical applications has been estimated to grow from \$1.4 billion in 2021 to \$2.4 billion by 2026 and the total global \acrshort{bci} market to \$6.2 billion by 2030~\cite{market-research, market-research-old}; consumer-neurotechnology research and development includes prominent companies such as Neuralink or Meta. A variety of full \acrshort{bci} systems as well as of wearable \acrshort{eeg}-headsets enabling \acrshort{bci} application development are now available on the market -- to what extent, however, all of these products truly record and process brain data as their main information source, as could be expected from a true \acrshort{bci} system, is not always entirely clear. Following the model of smartphones, neurotechnology companies provide developer tools such as \acrlong{sdk}s (\acrshort{sdk}s), \acrlong{api}s (\acrshort{api}s) for creating \acrshort{bci}-apps -- notably, in this way third-parties might gain access to collected brain data. There are companies encouraging and enabling users to share their own brain data recordings, e.g., with the promise to improve the \acrshort{bci}s' functionality~\cite{emotiv}. 

This rise of \acrshort{bci} technology has, over the past decade, also generated attention to new potential cybersecurity and privacy concerns. A fast-growing body of research literature addresses \acrshort{bci}-related privacy and security threats, risks, and their mitigation: As part of the current work we conducted an extensive search of the technical literature and identified 97 original research papers, topical reviews, and non-academic publications in the field of neuroprivacy and -security (see Chapter~\ref{chap:relatedwork}). The topic of neuroprivacy and -security has spurred debates in the general public, and, among other initiatives, fueled the worldwide NeuroRights campaign~\cite{neurorights}. In general, there is a broad consensus that neuroprivacy and -security are of crucial importance, and it has repeatedly been emphasized that both should be an integral part of the \acrshort{bci} design process (\verb|"security and privacy by design and default"|). But it is less clear 
\begin{itemize}
    \item how neuroprivacy and -security threats can be identified and the associated risks assessed,
    \item and how suitable mitigation strategies can be integrated into the \acrshort{bci} design 
    \item using transparent and systematic methodologies that can be reliably applied to the broad and heterogeneous spectrum of current and prospective \acrshort{bci} applications.
\end{itemize}

To  address these points, and to contribute to responsible research and innovation, our aim in the work presented here is to create a framework for neuroprivacy and -security which will be \textbf{generic} (suitable for different \acrshort{bci} types/paradigms in general, with necessary adjustments and amendments for specific cases), \textbf{use-case independent}, and \textbf{extensible}. The latter is important as neurotechnology is a highly dynamic, emerging field of innovation. Our framework and follow-up research aim to ensure and foster the understanding of the security and privacy challenges of current and future \acrshort{bci} usages as well as providing actionable guidelines for detection, evaluation, and mitigation of privacy- and security-related neurorisks. 

To this aim we will proceed in the following steps:
\begin{enumerate}
    \item We start out by providing a practical yet holistic working definition of \verb|"brain privacy"|, of the privacy-associated properties, and of the corresponding brain privacy threat categories; and we introduce the ideas of hard vs. soft privacy as well as of \verb|"privacy and security by design"|.  
    \item We then review the state and development of current \acrshort{bci} technology, propose an extended \acrshort{bci} cycle as the basis for our risk analysis, and derive a problem formulation on the background of the current \acrfull{trl} and prospective \acrshort{bci} technology.
    \item In the following \emph{~\nameref{chap:relatedwork}} section, we present an overview of the related work.  
    \item Next we introduce a basic neuroprivacy and -security architecture, and assess privacy threats using the {LINDDUN} privacy threat modeling methodology which has been developed to support analysts in systematically eliciting and mitigating privacy threats in software architectures, and estimate the risks associated with the identified threats~\cite{linddun}. As a result we identify  critical risks, such as in the categories linkability and content unawareness. 
    \item Applying generic privacy design patterns, we discuss and propose both general strategies and concrete mitigation techniques for all of the identified critical risks and map these solutions onto a generic privacy- and security-optimized \acrshort{bci} architecture. 
    \item In doing so, we also propose key research questions for the development of next-level privacy and security solutions for \acrshort{bci}s. 
    \item Along the way, we also address a series of more special topics, i.e., regional considerations in the European/German context, regulatory aspects, identification and authentication in the \acrshort{bci} context, how to ensure \acrshort{bci} user autonomy, as well as neurodata governance.

\end{enumerate}

%% file: chapters/2-definition.tex
\chapter{Brain Privacy}\label{chap:definiton}
\section{Working definition}
Privacy in general, and  \verb|"neuroprivacy"| or \verb|"brain privacy"| in particular, are not easy to define. For example, there have been multiple proposals to define neuroprivacy from a variety of fields such as neuroethics, neurosecurity, neuroscience, or neurolaw~\cite{bublitz_privacy_2019, gilead_can_2015, gligorov_brain_2016, ligthart_forensic_2021, minielly_privacy_2020, naufel_braincomputer_2020, richmond_i_2012, ienca_towards_2022, kellmeyer_big_2021}. General definitions of privacy vary and are contested across cultures, times, and contexts. For example, Tavani distinguishes unitary, derivative, and cluster definitions of privacy, as well as interest-based versus rights-based conceptions which have been adopted by philosophers and legal theorists, and notes that the meaning of privacy has considerably evolved, from relatively narrow concepts of privacy primarily related to property rights to much expanded notions of the right to privacy~\cite{himma-handbook-2008}. It appears unlikely that any single privacy definition will bring the debate about the nature and appropriate definition of privacy to an end; or that this is necessarily desirable given the diversity and ongoing evolution of privacy concepts. Thus, for the current work the aim was to select a practical yet holistic working definition of what is meant by \verb|"brain privacy"|. 

We base our working definition on the concept of informational privacy. Informational privacy is centered on access and control of personal information and can be defined as \emph{\say{having control over/limiting access to one’s personal information}}~\cite{himma-handbook-2008}. Informational privacy is thus different from the notion of physical privacy as non-intrusion with respect to one's physical space, decisional privacy as noninterference with respect to one's choices, and psychological/mental privacy as non-intrusion and/or non-interference with respect to one’s  thoughts and personal identity. It can be argued that the latter even adds to the problems of constructing a definition of privacy because a general definition of thoughts and mental states would be required in the first place; thus psychological/mental privacy notions may be useful to explain why privacy is valuable, but may be less helpful to define what privacy actually is~\cite{moore-information-privacy-rights}. Therefore, we base our working definition of brain privacy on the assumption that brain privacy involves a special case of informational privacy, namely, privacy with respect to information derived from brain data -- brain data in turn defined as any data obtained directly with respect to the brain, using a technical recording device~\cite{ienca_towards_2022}. 

This working definition of brain privacy as brain-data-related informational privacy has not only the advantage that it circumvents definition problems associated with terms as \verb|"mental states"| etc., but the concept of information appears well suited to connect across different disciplines relevant to the topic of neuroprivacy and -security. For example, our definition is compatible with the original definition of neuroprivacy as \emph{\say{privacy concerns with respect to mental and cerebral functioning as delineated through [information obtained by] brain imaging and other neurodiagnostic techniques}}~\cite{are-your-thoughts-your-own}. Information is a central concept in physical, neuro- and computer science. On the legal level, informational privacy is connected to the right to informational self-determination (Recht auf informationelle Selbstbestimmung\footnote[1]{ “[...] die Befugnis, grundsätzlich selbst zu entscheiden, wann und innerhalb welcher Grenzen persönliche Lebenssachverhalte offenbart werden” BVerfGE 103, 21 (33). Engl. “[...] the authority to decide for oneself when and within what limits personal life facts are to be disclosed”
}) in German law. Informational privacy has been analyzed to derive sets of privacy properties (Tab.~\ref{tab:table-1}), and these in turn correspond to privacy threats. For the analysis of privacy threats, there are systems engineering tools as we will discuss and apply later. 

\input{figures/2-definition/fig01-brain-privacy}

In light of our informational definition of brain privacy, it will also be important to consider what information exactly can be derived (is \verb|"decodable"|) from brain data, such as aspects of thoughts, motor intentions, or brain-health-related information, and how such data are generated in the physical brain. Reasons for this include that brain data may contain information that is not obvious to the \acrshort{bci} user, and attacks may target the level of neural signal generation~\cite{martinovic_feasibility_nodate, frank_using_2017}. We thus propose a multi-level approach, separately considering the Physical level (physical brain); the Data level (raw and processed brain data), in addition to the Decodable Information level as discussed above (including decoder model parameters, as they may also be informative). Fig.~\ref{fig:brain-privacy} shows these different levels of brain privacy map to a \acrshort{bci} cycle as introduced by van Gerven and colleagues~\cite{gerven_braincomputer_2009}. The \acrshort{bci} cycle is a helpful abstraction for analyzing \acrshort{bci} privacy and security issues~\cite{bernal_security_2022}.   

\section{Hard vs. soft privacy and the associated threat models}
Privacy can be broadly categorized as hard privacy and soft privacy, respectively~\cite{deng_privacy_threat_analysis}. The threat model of hard privacy assumes no trust in third parties. This includes organizational service providers (such as a \acrshort{bci} manufacturer in our case), data holders (e.g., a cloud storage service), as well as the general adversarial environment motivated to breach privacy; in our case for example hackers interested in stealing and selling brain data or secret information. 

In the soft privacy case, trust in third parties exists. The threat model is hence softer, including curious insiders, accidental data leaks, but also adversaries external to the trusted third parties. Unlinkability, anonymity \& pseudonymity, plausible deniability and non-detectability as in the {LINDDUN} methodology are hard privacy properties, content awareness and policy/consent compliance soft properties. Importantly, the hard and soft privacy scenarios imply different mitigation strategies. 

\begin{center}
\begin{table}[h!]
\centering
\begin{tabular}{ |c|c|c|c| } 
\hline
 &  & \textbf{Privacy properties} & \textbf{Threats}\\
\hline
\multirow{5}{5em}{LINDDUN} & \multirow{5}{3em}{Hard privacy} & unlinkability & linkability \\ 
\cline{3-4}
&& anonymity \& pseudonymity & identifiability\\ 
\cline{3-4}
&& plausible deniability & non-repudiation\tablefootnote[2]{Non-repudiation, in contrast to the security context, is considered a threat for privacy.} \\ 
\cline{3-4}
&& undetectability & detectability \\ 
\cline{3-4}
&& confidentiality & disclosure of information \\ 
\cline{2-4}
& \multirow{2}{3em}{Soft privacy} & content awareness & content unawareness \\
\cline{3-4}
&& policy/consent compliance & non-compliance \\ 
\hline
\multicolumn{2}{|l|}{\multirow{12}{8em}{\acrshort{gdpr}-derived privacy properties~\cite{huth-gdpr}}} & unlinkability & linkability  \\
\cline{3-4}
\multicolumn{2}{|l|}{} & pseudonymity /non-identifiability & identifiability \\
\cline{3-4}
\multicolumn{2}{|l|}{} & access control /authorization & uncontrolled access \\
\cline{3-4}
\multicolumn{2}{|l|}{} & integrity & data corruption \\
\cline{3-4}
\multicolumn{2}{|l|}{} & confidentiality & disclosure of information \\
\cline{3-4}
\multicolumn{2}{|l|}{} & availability/access & lack of access \\
\cline{3-4}
\multicolumn{2}{|l|}{} & data minimization & disproportionate data collection \\
\cline{3-4}
\multicolumn{2}{|l|}{} & information/transparency & content unawareness \\
\cline{3-4}
\multicolumn{2}{|l|}{} & storage limitation & disproportionate storage\\
\cline{3-4}
\multicolumn{2}{|l|}{} & purpose limitation & disproportionate processing\\
\cline{3-4}
\multicolumn{2}{|l|}{} & accountability & non-compliance\\
\cline{3-4}
\multicolumn{2}{|l|}{} & encryption & decryption\\
\hline

\end{tabular}
    \caption{Privacy properties which can be associated with our informational definition of brain privacy, in the LINDDUN privacy threat modeling methodology (top), and derived from the \acrshort{gdpr} (bottom), and as well as privacy threats associated with each of the privacy properties. Note that the two sets overlap but are not identical.}
    \label{tab:table-1}

\end{table}
\end{center}

\section{Privacy and security by design}
The phrases \verb|"privacy/security by design"| generally refer to making privacy and security part of the design process for new technologies and products, in contrast to \verb|"bolting on"| privacy and security functionality at a later (and, typically, too late) stage~\cite{gurses2011engineering}. The historical roots of privacy by design ideas can be traced back to the \acrfull{fipps} developed out of a 1973 report from the United States Department of Housing, Education, and Welfare (HEW report) ~\cite{waldman_gdpr_critic}. The term itself was coined later, together with the seven principles of privacy by design, Proactive not Reactive; Privacy as a Default Setting; Privacy Embedded into Design; Full Functionality; End-to-End Security; Visibility and Transparency; and Respect for User Privacy~\cite{cavoukian20107} 

European \acrfull{gdpr}, as in effect since 2018, now incorporates in Article 25 entitled \emph{\say{Data Protection by Design and by Default}}, the concept of \verb|"privacy by design"| into European data protection law. Given the early stage of \acrshort{bci} technology (see below and Fig.~\ref{fig:bci-trls}), in the current work we will put special focus on \verb|"by design"| strategies.

%% file: figures/2-definition/fig01-brain-privacy.tex
\begin{figure}[t]
\begin{centering}
    {\includegraphics[scale=0.36]{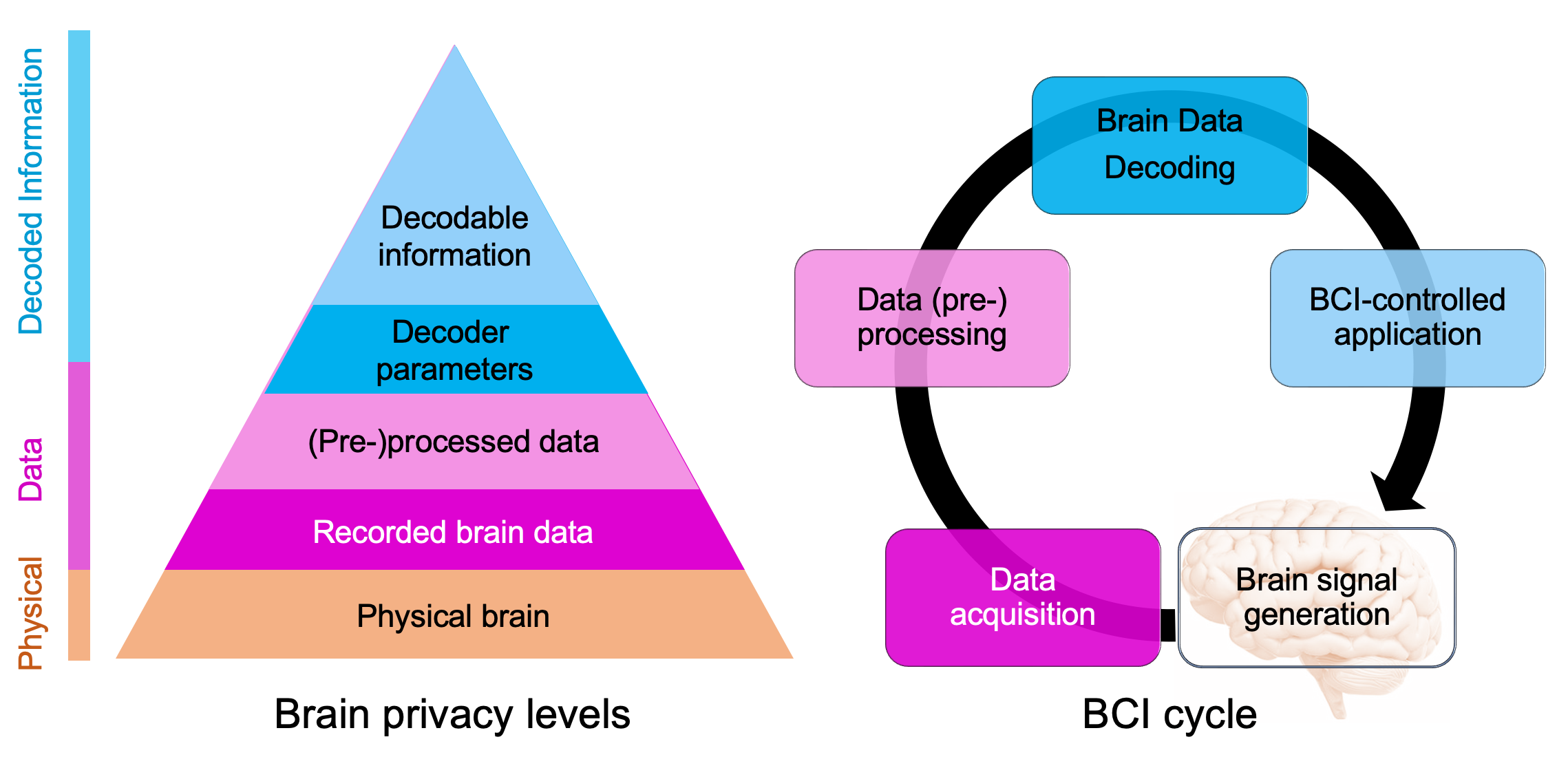}}
    \caption[Multi-level approach to brain privacy]{\textbf{Multi-level approach to brain privacy (left).} The different levels of description relevant for a definition of brain privacy are mapped by corresponding colors to the components of the core \acrshort{bci} cycle~\cite{gerven_braincomputer_2009} (\textbf{right}). This mapping will serve as a bridge between the conceptual definition of brain privacy on the one side, and the implementation of a basic safety architecture on the other.}%
    \label{fig:brain-privacy}
\end{centering}
\end{figure}

%% file: chapters/3-bci-concepts.tex
\chapter{BCI Concepts and Technology}\label{chap:bci-concepts}
\section{BCI taxonomy}

A \acrshort{bci} has been defined as a system \emph{\say{that measures \acrfull{cns} activity and converts it into artificial output that replaces, restores, enhances, supplements, or improves natural CNS output and thereby changes the ongoing interactions between the CNS and its external or internal environment}}~\cite{wolpaw_braincomputer_2012}. From its beginnings in the 1960s and 70s, \acrshort{bci} technology has now developed into a broad and multifaceted research area and technology class, spanning fields from assistive devices for paralyzed patients to neurogaming, from industry and work-related productivity tools to education, and from military to artistic applications. \acrshort{bci}s may range from high-end research prototypes to low-cost consumer-grade commercially-available products. 

The growth of the \acrshort{bci}s domain is reflected in the fact that it can be structured along multiple (though partially related) dimensions: 
\begin{enumerate}
    \item Direction of information flow: Recording, stimulating (unidirectional \acrshort{bci}s), or both (bidirectional \acrshort{bci}s). 
    \item Invasiveness: Invasive, minimally / semi-invasive\footnote[4]{The definition of minimally and semi-invasive BCIs is not uncontroversial. Semi-invasive BCIs can, e.g., be defined as those which “involve recording from or stimulating the brain surface or nerves.” Rao, R. P. (2013). Brain-computer interfacing: an introduction. Cambridge University Press, p. 101). Others consider, e.g.,  BCIs based on clinical ECoG grids, as more rather than less invasive compared to intracortical BCIs utilizing small implants like the Blackrock-Array. }, non-invasive; 
    \item Relatedly, recording techniques: Scalp \acrfull{eeg}, \acrfull{meg}, \acrfull{fnirs}, functional transcranial Doppler ultrasound, \acrfull{fmri}, subcutaneous \acrshort{eeg}, skull screws, epidural recordings, \acrfull{ecog}, micro-\acrshort{ecog}, stereo-\acrshort{eeg}, intracortical \acrfull{lfp}, \acrfull{sua} or \acrfull{mua}; 
    \item Type of control: Active, passive, reactive; 
    \item Timing of information extraction: Synchronous and asynchronous; 
    \item \verb|"Non-hybrid/unimodal"| vs. \verb|"bi-/multimodal/hybrid"| \acrshort{bci}s; 
    \item Medical vs. non-medical \acrshort{bci}s with corresponding regulatory consequences; 
    \item Relatedly, the interfacing target: brain to robot, to internet (\emph{\say{Internet of Neurons}}~\cite{sempreboni_privacy_2018}), to vehicle~\cite{nissan-bitbrain, daimler-nextmind}, silent communication/\verb|"Synthetic Telepathy"|~\cite{brigham_imagined_2010}, smart-home control, brain-to-spine and brain-to-brain interfaces, both to human and animal~\cite{zhang_human_2019}, etc.; 
    \item \acrshort{bci} paradigm / neurophysiological principle: \acrshort{bci}s based on operant conditioning, based on population decoding, slow-cortical-potential-based, P300 spellers, learned control over oscillatory brain signals; motor-imagery-based, steady-state sensory evoked potentials, e.g., visual: \acrlong{ssvep}s (\acrshort{ssvep}s); 
    \item Type of brain signal processing/decoding: with and without \acrfull{ml}; with \acrshort{ml}: linear vs. nonlinear classifiers; traditional (\acrfull{svm}, \acrfull{rlda}, random forests etc.) vs. \acrfull{dl}-based; 
    \item Single-purpose vs. multi/general-purpose \acrshort{bci}s; 
    \item Single user vs. multi-user (collaborative or competitive) \acrshort{bci}s. 
\end{enumerate}

For example, the NextMind \acrshort{bci} could be classified as a non-invasive, reactive, synchronous, non-hybrid, consumer single user multi-purpose \acrshort{ssvep}-based \acrshort{bci} using an undisclosed \acrshort{ml} algorithm~\cite{nextmind}.

A special type of \acrshort{bci}s uses brain signals (mostly \acrshort{eeg}) as a biometrics for both user identification and authentication. This growing field of research has been extensively reviewed by~\cite{jalaly_bidgoly_survey_2020, chan_challenges_2018} highlighting that brain signals generally comply well with the general desiderata of biometric signals\footnote[5]{Universality, distinctiveness, collectability, circumvention, permanence, acceptability and performance~\cite{chan_challenges_2018}}. As an example, it can be argued that \acrshort{eeg} has superior universality compared to retina scanning or fingerprints. Beyond its potential as a general biometric modality, brain-signal-based biometrics such as using the \acrshort{eeg} may be used for authentication and identification of users of \acrshort{bci} applications~\cite{boubakeur_eeg-based_2017}. Such approaches could leverage specific brain responses that are generated during \acrshort{bci} usage anyhow, and thus enable seamless authentication with minimal additional burden on the user. This setup would be a \acrshort{bci} (for biometrics) within a \acrshort{bci} (for another purpose), illustrating the growing complexity of contemporary \acrshort{bci} setups (posing a challenge for coming up with comprehensive \acrshort{bci} taxonomies). Next we consider the hard- and software components used to implement these diverse types of \acrshort{bci}s.

\section{BCI hard- and software}
The scope of hard- and software used in \acrshort{bci}s is broad, with important consequences for neurosecurity and -safety. \acrshort{bci}s may be implemented either using off-the-shelf components or custom hardware, such as custom \acrlong{asic}s (\acrshort{asic}s). Embedded \acrshort{bci}s have been realized, e.g., using the Raspberry Pi and Arduino platforms; future embedded \acrshort{bci} solutions might utilize neuromorphic computing hardware. 

Many \acrshort{bci}s (but not all) involve a wireless connection to a local computing device for data processing and storage; and may involve remote/cloud services and components. Such \acrshort{bci}s may also utilize mobile phones and other general-purpose hardware. Presence or absence of wireless communication or trusted hardware may result in very different conditions for the establishment of effective \acrshort{bci} cybersecurity. 

\begin{figure}[h!]
\begin{centering}
    {\includegraphics[scale=0.4]{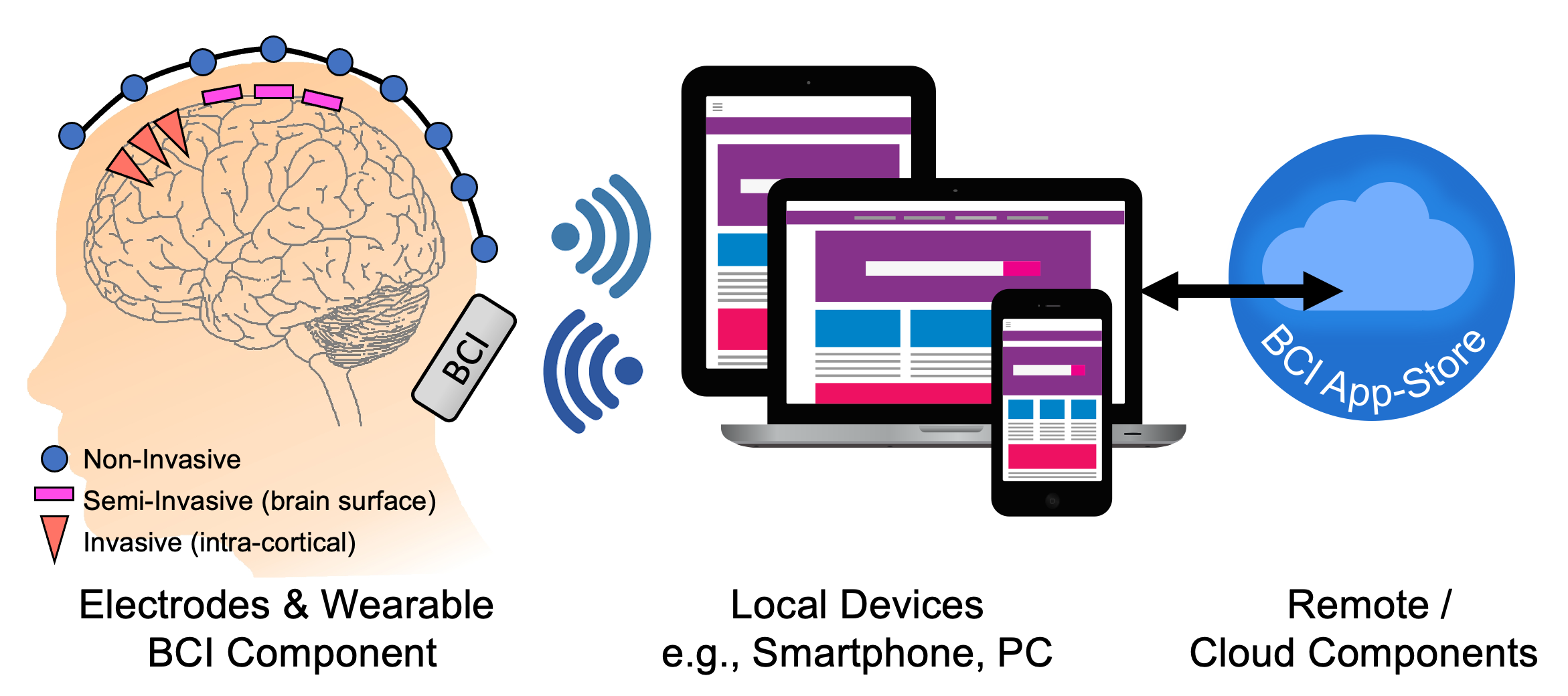}}
    \caption[BCI hardware overall structure]{\textbf{BCI hardware overall structure} may involve wearable/embedded components (left), components running on local devices (center), as well as remote/cloud services (right). }%
    \label{fig:hardware-software}
\end{centering}
\end{figure}

In the one extreme, all \acrshort{bci} functions may be on the wearable component (so there is no local or cloud component); in the other extreme, all functions other than recording (and stimulation) may be supported by cloud services. In the typical case, at least as currently prevailing, a substantial part of the \acrshort{bci} functionality is supported by a local device, which may be coupled wirelessly (e.g., Wi-Fi, Bluetooth) or via a wire connection to those components directly connected to the \acrshort{bci} user. 

The wide spectrum of \acrshort{bci} hardware is paralleled on the software side. In addition to the diversity of specific \acrshort{ml} components used in \acrshort{bci}s as mentioned above, there are different general \acrshort{ml} frameworks relevant to \acrshort{bci}s, such as PyTorch and TensorFlow for \acrshort{dl}, as well as numerous \acrshort{bci} software frameworks using different programming languages, e.g., Open BCI GUI, Open ViBe, Python MNE (using Java, Python, C{++}, NodeJS) and Letswave, EEGLab, BCILab (Matlab toolboxes). 

\section{The extended BCI cycle}
In light of this conceptual and technological diversity it is crucial to identify generic \emph{functional principles} which are shared by most or even all (currently conceivable) \acrshort{bci}s. One such principle has been formalized as the so-called \textbf{\acrshort{bci} cycle}. It was originally described as a closed loop, sequentially involving the measurement of brain activity, classification of the recorded data data, the feedback to the subject as well as the resulting effect back on brain activity of the \acrshort{bci} subject. 

In the subsequent literature, several modifications and extensions of the \acrshort{bci} cycle have been put forward. In particular with respect to the \acrshort{bci} cycle as a basis for evaluating cybersecurity, Bernal and colleagues proposed a version of the \acrshort{bci} cycle including stimulation functionality~\cite{bernal_security_2022}. However, as our focus here is not on bidirectional \acrshort{bci}s with stimulation functions, and as current unidirectional \acrshort{bci}s without stimulation functions have developed far beyond the initial, comparable simple systems, here we propose a different extension of the \acrshort{bci} cycle (Fig.~\ref{fig:bci-cycle}). This extended \acrshort{bci} cycle is designed to accommodate the functionality of \textbf{current and prospective “three level” \acrshort{bci} setups}:
\begin{enumerate}
    \item The \textbf{\acrshort{bci} Core Cycle} corresponds to the classical \acrshort{bci} cycle and interacts with a single user (or a group of users in the case of a collaborative \acrshort{bci}). 
    \item The \textbf{Extended Core} may comprise various possible extensions of the core functionality, which may communicate with the core, the global components, or both. There are various such existing or prospective modules extending the core functionality, e.g., modules to implement adaptivity of the core decoder, \acrshort{bci} app store client modules, extension modules for anomaly detection with respect to the core data flow, etc.
    \item The \acrshort{bci} \textbf{Global Functionality}. The main distinction to the (extended) core is that the global functions concern multiple (other) users, e.g., data pooling across multiple users, training global models on such pooled data, or \acrshort{bci} app store servers.
\end{enumerate}
Many different scenarios are possible of how this functionality may map onto the wearable/embedded, local, and remote/cloud components of a \acrshort{bci}. For example, the core functions may be fully implemented in the embedded/wearable component, and extensions as described above would fit well to a local device such as a PC or smartphone. On the other hand, decoders may also be run locally, on a remote server (such as done in~\cite{kuhner_service_2019}), or even as a cloud service (provided appropriate stability and latency of the remote connection for the given application, such as cloud neurogaming). 

\include{figures/5-bci-concepts/fig03-bci-cycle}

\section{Current state and future trajectory of BCI technology}

\acrfull{wipo} Technology Trends 2021 report on assistive technology, analyzing patenting and technology trends in assistive technology, using a scale of technology readiness. The report estimates the typical readiness of \acrshort{bci} technology as comparably low, somewhere between  \verb|"roof of concept"| to a \verb|"minimal viable product"|~\cite{world_intellectual_property_organization_wipo_0}. Many intriguing \acrshort{bci} application concepts are still in the stage of (academic) lab prototypes. Many of those still are still fraught with fundamental issues which need to be resolved for viable \acrshort{bci} products. One bottleneck, for example, is the limited amount of information which can be derived from recordable brain signals, especially in the non-invasive cases. 

This limitation may be overcome by novel recording techniques, such as \acrshort{meg} with \acrfull{opm}. Also progress in the field of \acrshort{ml}, especially in the field of \acrshort{dl} with \acrfull{ann}, can significantly increase the amount of extractable information from various data types. \acrshort{dl} has set a new state of the art in computer vision and natural language processing. Despite promising results, \acrshort{dl} has not yet enabled similarly large performance increases in brain signal decoding. One possible reason for this difference lies in the lack of \acrshort{bci}-related \verb|"big data"|. Thus, data pooling such as \acrshort{eeg} recordings may be crucial to move the \acrshort{bci} field from lab prototypes to real-world products. In other words, current \acrshort{bci} applications typically do not have a lot (if any) \verb|"spare performance"| that could be sacrificed for improved security or privacy, in such cases where security- and privacy-preserving methods come at a performance cost (see below). 

\begin{figure}[h!]
\begin{centering}
    {\includegraphics[scale=0.4]{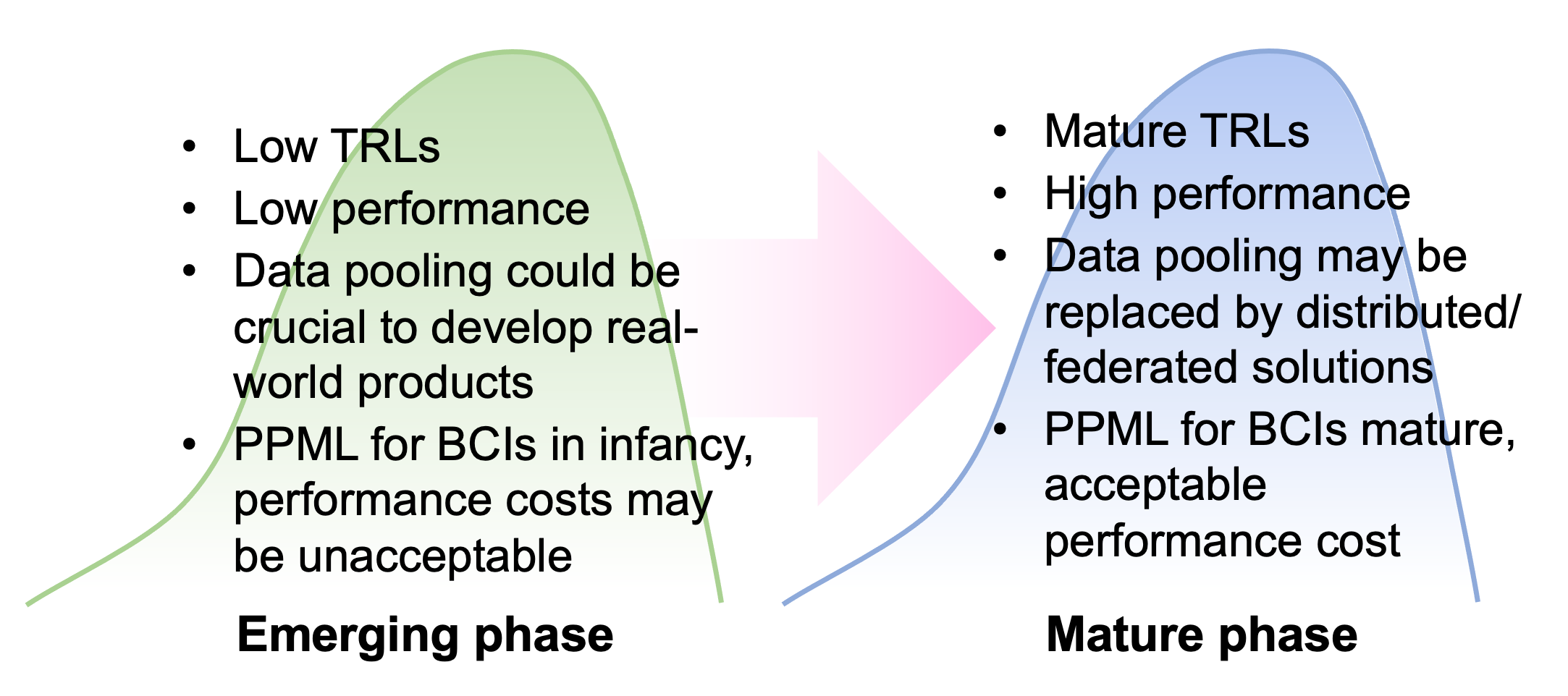}}
    \caption[Current and anticipated future phases in the maturation of BCI technology]{\textbf{Current and anticipated future phases in the maturation of BCI technology.} Privacy and security solutions should match the stage of development of BCIs and thus face evolving challenges. 
}%
    \label{fig:bci-trls}
\end{centering}
\end{figure}

\acrshort{bci} technology today can be seen in an \verb|"emerging technology phase"|, characterized by relatively low \acrshort{trl}s, little or no spare performance (decoding accuracy), by the fact that large-scale data pooling could be crucial to develop better \acrshort{ml} models and real-world products, and that \acrfull{ppml} for \acrshort{bci}s is still in a research stage and technological infancy; when \acrshort{ppml} incurs a performance cost  this may be problematical given the limited amount of decodable information \acrshort{bci}s have to work with anyhow. 

We anticipate that after a transition phase, \acrshort{bci} technology will enter a \verb|"Mature Technology Phase"|, characterized by functional \acrshort{trl}s/real-world products, based on solid decoding performance, a situation where data pooling may be replaced by distributed/federated \acrshort{ml} solutions. 

Considering this technology trajectory is crucial in our context, as it implies three distinct strategic objectives:
\begin{enumerate}
    \item \textbf{Act}: Come up with concrete neurosecurity and -privacy recommendations which are already practical \verb|"here and now"|, during the ongoing emerging technology phase of \acrshort{bci}s.
    \item \textbf{Research}: Identify key neurosecurity and -privacy research questions to be addressed during the transition phase.
    \item \textbf{Anticipate}: Project how neurosecurity and -privacy issues may be solved in the anticipated mature \acrshort{bci} technology phase. 
\end{enumerate}

%% file: figures/5-bci-concepts/fig03-bci-cycle.tex
\begin{figure}[t]
\begin{centering}
    {\includegraphics[scale=0.4]{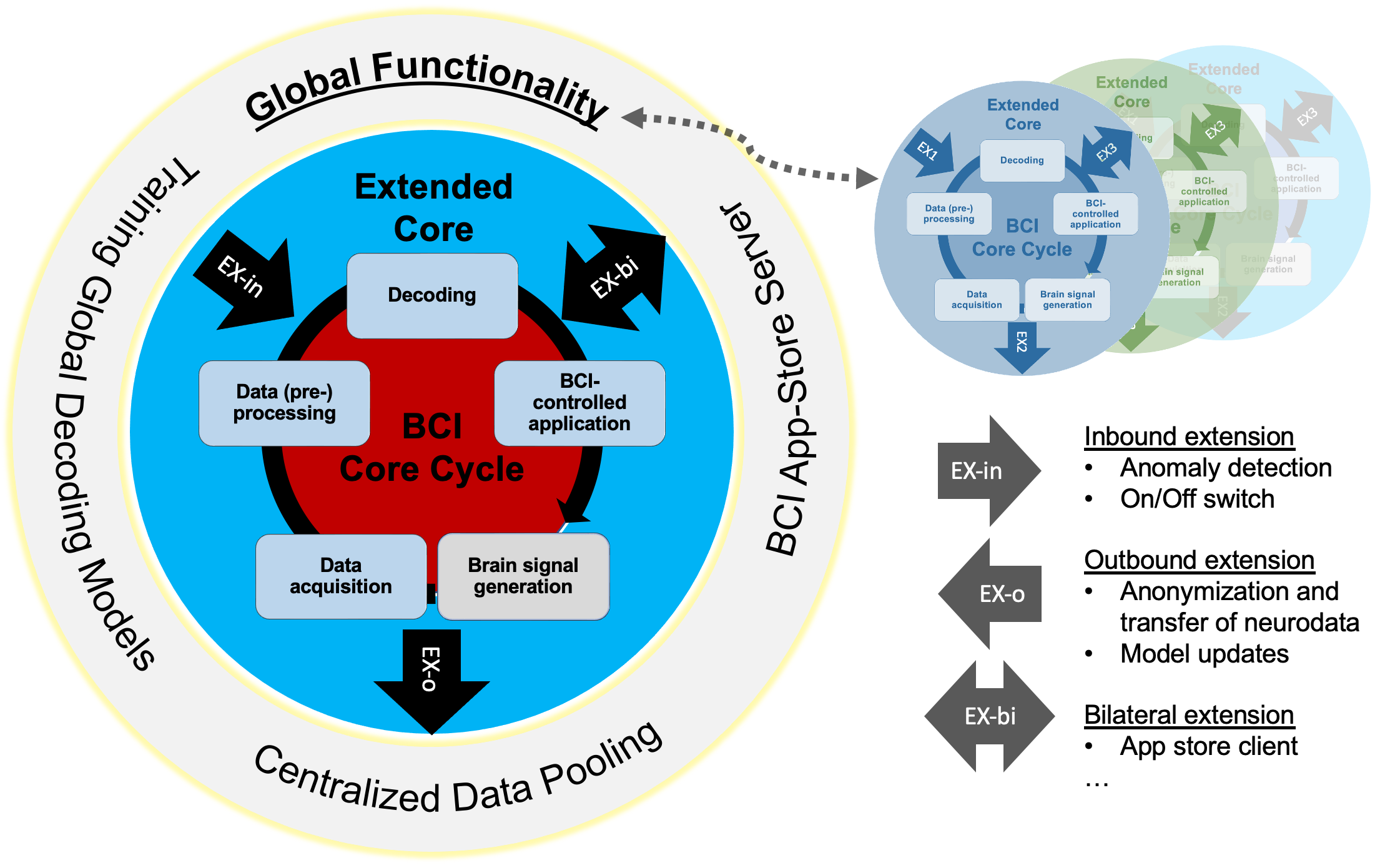}}
    \caption[Extended BCI cycle]{\textbf{Three-shell, extended BCI cycle.} The \textbf{BCI Core Cycle} corresponds to the classical BCI cycle and interacts with a single user (or a group of users in the case of a collaborative BCI). The \textbf{Extended Core} may comprise various possible extensions of the core functionality, which may communicate with the core, the global components, or both. The BCI \textbf{Global Functionality} is defined with respect to multiple cores of other users (symbolized by colored pictograms on the top right), e.g., data pooling across multiple users, training global models on such pooled data. etc. - note that this distinction is based on function, not on hardware implementation; there are numerous different possibilities how this generic functionality can be mapped on hardware (cf.~\ref{fig:hardware-software}). This three-shell BCI cycle will be used to derive a data flow diagram, which in turn is the basis for the subsequent systematic privacy threat modeling.  }%
    \label{fig:bci-cycle}
\end{centering}
\end{figure}

%% file: chapters/4-relatedwork.tex
\chapter{Related Work}\label{chap:relatedwork}
\begin{figure}[h!]
\begin{centering}
    {\includegraphics[scale=0.4]{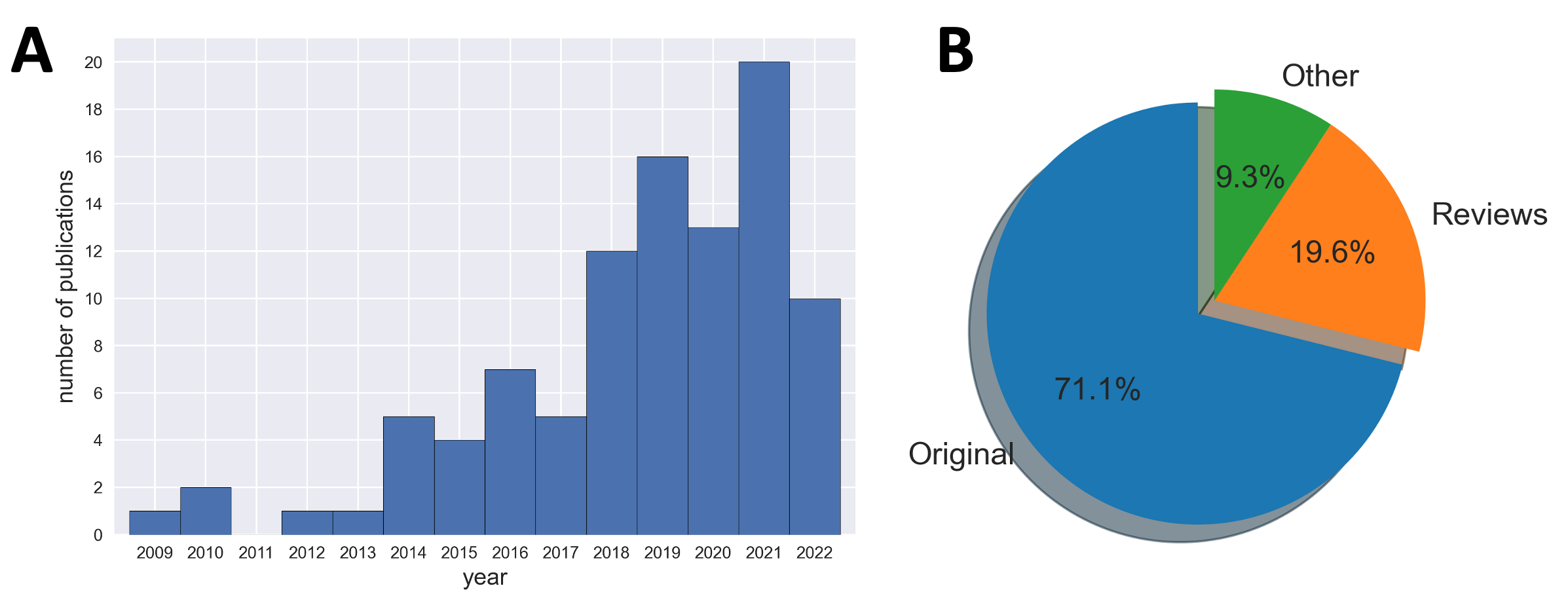}}
    \caption[Neuroprivacy and -security literature overview]{\textbf{A} Publications on technical aspects of neuroprivacy and -security from 2009 until June 2022; \textbf{B} pie chart of paper types; other: editorial, opinion paper, blog post, non-peer-reviewed article, PhD thesis, or book chapter. 
}%
    \label{fig:related-work}
\end{centering}
\end{figure}

We performed a literature search and identified 97 publications on technical aspects of \acrshort{bci} neurosecurity and -privacy (Fig.~\ref{fig:related-work}). Around 15\% of original studies (11 papers) actually implemented attacks on \acrshort{bci}s and demonstrated successful cyberattacks on \acrshort{bci}s in lab or more real-world settings. Some of the studies were done in laboratory settings, such as Belkacem et al. who implemented \acrfull{ddos} and Man in the Middle attacks on P300-based \acrshort{bci} Unicorn speller~\cite{belkacem_cybersecurity_2020}; Beltran and colleagues successfully implemented misleading stimuli attacks (see below) and different versions of noise-based attacks on P300-based \acrshort{bci}s~\cite{cyberbrain, martinez_beltran_noise-based_2022, gupta_secbrain_2022}. Interestingly, the majority of the implemented and emulated attacks were performed on \acrfull{erp}-based reactive \acrshort{bci}s. 

One of the first documented successful hacking attempts was Cody’s Emokit project, where Cody Brocious cracked encrypted data directly from EMOTIV headset~\cite{EMOKIT}. Later on, Cusack and colleagues used the \emph{\say{Btlejuice}}\footnote[6]{Btlejuice https://github.com/DigitalSecurity/btlejuice} attack: \acrshort{eeg} data were hijacked between Emotiv headset and \acrshort{ncd} by tethering in Bluetooth wireless network~\cite{cusack_neurosecurity_2017}. Similarly, Sundararajan showcased several attacks on Emotiv headset, such as passive eavesdropping, active interception and \acrshort{ddos} (by jamming the connection from Emotiv to smartphone)~\cite{sundararajan}.  Xiao and colleagues implemented  two sets of \acrfull{poc} attacks, consisting of four remote and one proximate attack. Results showed that all 156 \acrshort{bci} apps in the Neurosky app store are vulnerable to the proximate attack and all the 31 free apps are vulnerable to at least one remote attack~\cite{xiao_i_2019}. Finally, another recent large scale study by Tarkhani et al. used the system security and privacy threats analysis for existing wearable \acrshort{bci} products (Muse, NeuroSky and OpenBCI) from the operating system and adversarial \acrshort{ml} perspectives. They designed an information flow control system for attack mitigation and as \acrshort{poc} they implemented a set of attacks across six vectors (\acrshort{av}1: sniffing, spoofing, man in the middle; \acrshort{av}2: inadequate isolation and access control; \acrshort{av}3: privilege escalation; \acrshort{av}4-\acrshort{av}6: adversarial \acrshort{ml}) and discovered more than 300 vulnerabilities for real-world \acrshort{bci} devices~\cite{tarkhani_enhancing_2022}. For an overview of cybersecurity threats taxonomy for \acrshort{av}s see Bernal et al.~\cite{bernal_security_2022}. 

Together, these studies clearly show that cybersecurity issues are an important concern for existing consumer neurotechnology; therefore state-of-the-art cybersecurity defense strategies are of great importance for \acrshort{bci}s. At the same time, these studies also highlight that \verb|"hacking the BCI"| is a different, yet related, topic area compared to \verb|"hacking the brain"| violating brain privacy (the latter being the focus of our work described here). It is thus important to distinguish between the following different, but tightly entangled aspects of \acrshort{bci} security: Using \verb|"conventional"| hacking techniques to get access to a \acrshort{bci} system. Here, the defense is provided by \verb|"conventional"| cybersecurity measures. Using the \acrshort{bci} to \verb|"hack the BCI"| - e.g., to steal secret/personal information, or to compromise the autonomy of the \acrshort{bci} user. Here, the defense lies in the design of the \acrshort{bci}s. From our literature review here, however, few papers clearly distinguish between the two concepts of \verb|"hacking the BCI"| vs \verb|"hacking the brain"|.

A widely explored topic in the literature is misleading stimuli attacks, or so-called \verb|"brain malware"|. The most common ways of using malicious stimuli to extract private information are oddball paradigms, guilty knowledge tests, and priming. For example, Vliet and colleagues showed that the N400 \acrshort{erp} component can be used to determine what a user is primed on~\cite{vliet_guessing_2010}. Other studies used consumer grade headset and showed the feasibility of detection of subconscious face recognition from \acrshort{erp}s~\cite{martin_detection_2016, bellman_wksp_2018}. Such methods can potentially also be used to infer other subconscious information such as implicit associations, preferences etc. For the excellent review of detection of concealed information from the P300 as well as possible deception strategies, see Rosenfeld~\cite{rosenfeld_p300_2011}. The first study which showed the feasibility of such an attack was already reported in 2012 by Martinovic and colleagues~\cite{martinovic_feasibility_nodate}. Utilizing the P300 paradigm and using different types of images, they demonstrated the feasibility to infer PIN codes, bank information, the month of birth, familiar faces, and geographical locations of the user. The results were further reproduced and extended by Lange et al.~\cite{lange_side-channel_2017}. Using a \emph{\say{Flappy Whale}} \acrshort{bci} game, Bonaci reported the feasibility of probing private information even with subliminal (not consciously perceived) stimuli~\cite{bonaci_phd_2015}, similar to Frank et al., who also reported that by analyzing the responses evoked by short stimuli hidden in video frames it was possible to uncover whether participants were familiar with the subliminally-presented faces~\cite{frank_using_2017}. A number of more recent studies, however, tried unsuccessfully to reproduce results with subliminal stimuli~\cite{martinez_beltran_noise-based_2022, quiles_perez_breaching_2021}. 

Among the proposed countermeasures for misleading stimuli attacks, the most prominent and frequently mentioned is the \verb|"BCI Anonymizer"| for which a patent was filed (but later abandoned) in 2014~\cite{bci-anonymizer}. To our knowledge, however, it still has not been successfully implemented even in laboratory settings due to lack of neuroscientific understanding of the \acrshort{eeg} signal and corresponding difficulties in filtering out task-irrelevant identifying information. Other measures include concentration on non-target stimuli, adding noise to raw \acrshort{eeg} data and \acrshort{api} restriction (non-exposing raw \acrshort{eeg} data to third parties). However, the effectiveness of the countermeasures (except for \acrshort{api} restriction) still requires further research. 

Despite the significant number of reviews present in our search, systematic privacy and/or security threat modeling is very scarce. Bonaci~\cite{bonaci_phd_2015} used an approach developed by Friedman for \emph{\say{Value Sensitive Design}}~\cite{doorn_value_2013, hutchison_development_2006}. She identified the following general threat categories for \acrshort{bci} technologies: 1. Disclosure to Unauthorized Parties, 2. Unauthorized Use of Individual Data, 3. Unauthorized Request (Search) for Individual Data, 4. Unauthorized Use of Aggregated Data, 5. Unauthorized Fusion of \acrshort{bci} data with Unexpected External Data. Another attempt at systematically evaluating privacy in \acrshort{bci}s was done by Wahlstrom and colleagues~\cite{wahlstrom_privacy_2016}. This study discusses potential privacy disruptions for \acrshort{bci} typology (active, passive, reactive, hybrid) and for existing, prospective and speculative use-cases. Different privacy theories (control, restricted access, commodification, contextual and ontological) were used for analysis. The result indicates that while all four types of \acrshort{bci}s have potential for disrupting privacy, the major risk is likely to arise from the use of reactive, passive and hybrid \acrshort{bci}s. Pazouki and colleagues attempted to use STRIDE~\cite{stride} for identifying \acrshort{bci} security risks, but the paper is lacking in detail on this topic~\cite{pazouki_false_2021}. 

\begin{tcolorbox}[colback=RoyalBlue!5!white,colframe=RoyalBlue!75!white]
\textbf{Literature survey conclusions}: 
\begin{itemize}
    \item There are relatively few publications addressing \acrshort{bci} security and privacy by design, and few publications clearly distinguishing between \verb|"hacking the BCI"| and \verb|"hacking the brain"|, and, to our knowledge, no publications applying a combination of systematic threat modeling, risk assessment, and privacy or security engineering tools to the field of \acrshort{bci} technology.
    \item Safety and privacy by \acrshort{bci} design to prevent \verb|"hacking the brain"| in \acrshort{bci}s is the central topic of the proposed framework.
\end{itemize}
\end{tcolorbox}

%% file: chapters/5-architecture.tex
\chapter{Basic Neuroprivacy and -security Architecture}\label{chap:architecture}
\section{Methodology overview}
To establish a basic neuroprivacy and -security architecture in a systematic manner, here  we leverage systems engineering methodologies. \textbf{Privacy engineering} addresses privacy issues in a systematic and holistic way and provides patterns on how to move from principles to actions. Privacy engineering has been defined as \emph{\say{a specialty discipline of systems engineering focused on achieving freedom from conditions that can create problems for individuals with unacceptable consequences that arise from the system}} as it processes personal information~\cite{brooks_introduction_2017}.

\input{figures/7-architecture/fig06-flowchart}

In the first step, we derived a \acrfull{dfd} based on our functional model (extended \acrshort{bci} cycle, Fig.~\ref{fig:bci-cycle}) and overall hardware structure (Fig.~\ref{fig:hardware-software}). The following steps then involve privacy threat modeling, risk assessment, and privacy engineering based on a set of basic privacy design patterns (Fig.~\ref{fig:flow-chart}).  

\section{Privacy threat modeling}\label{sec:privacy_threat}
To establish a basic neuroprivacy and -security architecture, we start with a systematic approach to modeling of brain privacy threats. As shown by our literature review above, such a systematic evaluation is so far lacking. We selected LINDDUN as our threat modeling tool. Similar to Microsoft’s threat modeling framework STRIDE, LINDDUN leverages an information-flow-oriented system model to systematically guide the threat analysis and to provide a broad coverage of different threat classes. In contrast to STRIDE, LINDDUN is oriented towards privacy threats and thus appeared especially well suited, given the pivotal role of brain privacy in our context\footnote[7]{However, we are well aware that these threat modeling methodologies were not designed for the very specific cases of brain privacy and interfacing; thus we considered it as a question to be answered whether or not this  approach would turn out to be useful in the BCI context. }. Furthermore, LINDDUN can be used not only for single systems but also for broader analyses; for example Iwaya et al. recently used LINDDUN to analyze a large group of top-ranked mental health apps from Google Play Store~\cite{iwaya_privacy_2022}.

Three \textbf{threat sources} are distinguished in LINDDUN: \emph{organizational}, which may refer to the organization as a whole or an employee, in our case first of all the \acrshort{bci} service provider; \emph{external}, which refers to a misactor external to the organization, in our case for example a malicious adversary with a motivation to steal brain data or sensitive information derived from it; and \emph{(future) receiving parties}: e.g., a cloud storage service, \acrfull{aws}, PayPal, etc., receiving data from the organization.

Threat modeling according to LINDDUN proceeds in three steps: 
\begin{enumerate}
    \item Modeling the system using a \acrshort{dfd}
    \item Eliciting threats using the LINDDUN threat catalog and mapping the identified threats to hotspots in the DFD
    \item Identifying mitigation strategies. 
\end{enumerate}
Thus, as the first step, we derived a \acrshort{dfd} from the extended \acrshort{bci} cycle schematic as introduced and discussed in the previous sections. The resulting \acrshort{dfd} is shown in Fig.~\ref{fig:dfd}.

\input{figures/7-architecture/fig07-dfd}

Next, we systematically evaluated the 34 threats in the 7 threat classes: linkability, identifiability, non-repudiation, detectability, disclosure of information, content unawareness, and noncompliance, as provided by the LIND(D)UN GO\footnote[8]{Note that the second “D” in LINDDUN stands for “Disclosure of information” which is a security category. It is not included in LINDDUN GO as it focuses on privacy. The LINDDUN authors however advise combining their approach with security threat modeling as privacy highly depends on security. For the same reason, in our framework we will also consider the relevant security aspects, such as with respect to disclosure of information and confidentiality.} system~\cite{WuytsKim2020LGAL}. In addition to the standard evaluation, we also asked whether each of the threats may be considered \acrshort{bci}-specific.  The LINDDUN framework (as also STRIDE) supports threat but not risk assessment and is compatible with the standard risk assessment techniques. We chose the widely used \acrfull{owasp}’s Risk Rating Methodology to estimate the risk associated with each identified \acrshort{bci}-specific threat. This means, to keep the document concise, \textbf{here we focused on the clearly \acrshort{bci}-specific threats, as opposed to the less- or non-\acrshort{bci}-specific threats} (however, it would be straightforward to extend the same kind of analysis also to the latter). \acrshort{owasp} overall risk severity ranges from \verb|"note"|, \verb|"low"|, \verb|"medium"| and \verb|"high"| to \verb|"critical"| (critical risk level for threats with high associated impact and likelihood)~\cite{owasp}. 

The aims when applying these methodologies here (i.e., to a generic model representing the architectures across the wide, emerging field of \acrshort{bci} technologies) obviously must be different from the aims when applying these methodologies to an existing and concrete system. Existing and prospective \acrshort{bci}s, e.g., range from low-cost systems with a few, low-quality \acrshort{eeg} sensors to high-end \acrshort{bci}s with potentially thousands of implanted electrodes. Accordingly, the amount of personal information that can be inferred from the recorded brain data acquired with different kinds of \acrshort{bci}s (low- vs. high-end, non-invasive vs. invasive) can be expected to vary enormously. As a consequence, identical privacy threats can have very different impacts and, consequently, lead to very different privacy-related risks. 

Therefore, our aim is an assessment of the landscape of potential \acrshort{bci}-specific privacy threats and risks. Thanks to the systematicity of the system engineering tools applied, it is also the aim to lay out a \textbf{general and extensible framework that can be flexibly adapted and applied to concrete and specific \acrshort{bci} application examples.} It was not our aim to provide ready-made/out-of-the-box risk assessments and mitigation recipes which can be just copied and put onto arbitrary \acrshort{bci} application scenarios. 

\begin{tcolorbox}[colback=TealBlue!5!white,colframe=TealBlue!75!white]
\textbf{LINKABILITY threats}: 
\begin{itemize}
    \item L1 -- Linkability of credentials (Organizational)
    \item L2 -- Linkable user actions (Organizational)
    \item L3 -- Linkability of inbound data (Organizational)
    \item L4 -- Linkability of context (Organizational/External)
    \item L5 -- Linkability of shared data (Receiving party)
    \item L6 -- Linkability of stored data (Organizational)
    \item L7 -- Linkability of retrieved data (Organizational, [Future] receiving party)
\end{itemize}
\end{tcolorbox}

The first threat category in the LINDDUN catalog comprises \textbf{LINKABILITY threats}, threat L1 being LINKABILITY OF CREDENTIALS: Actions and data can be linked by re-using credentials (for multiple system interactions). We considered this risk as non-\acrshort{bci} specific and therefore did not include it in our further analysis. The first \acrshort{bci}-specific threat which we identified was L3 LINKABILITY OF INBOUND DATA (The data sent to the system are linked to already collected data of the same or other data subjects, from the same or other source). We considered this threat as \acrshort{bci}-specific because correlating different brain signal recordings of the same subject in a data-driven manner is a highly specific task due to the special signal properties of such data, e.g., \acrshort{eeg} recordings. However, we also argue that the main risk from such data linkage arises in the stage of stored data (especially for remote/cloud storage of neuro-data), thus we assigned only a medium risk to item L3, but critical risk level to L6 LINKABILITY OF STORED DATA. Stored data are not only the raw or processed brain signals, but also the context, such as visual and auditory stimuli presented (and possibly also other modalities such as electrooculo- or cardiogram). Without knowing the context of \verb|"what happened"| with precise synchronization, the information that can be extracted from brain signals is reduced. For example, the whole class of misleading stimuli attacks via \acrshort{bci}s requires exact knowledge of when the misleading stimuli appear in the data stream. Thus, we classify linkability of stored \acrshort{bci}-acquired data as a critical risk. The same reasoning also applies to L7 LINKABILITY OF RETRIEVED DATA, provided the scenario that data would be passed on to third, receiving parties.

\begin{tcolorbox}[colback=TealBlue!5!white,colframe=TealBlue!75!white]
\textbf{IDENTIFIABILITY threats}: 
\begin{itemize}
    \item I1 -- Identifying credentials (Organizational)
    \item I2 -- Actions identify users (Organizational)
    \item I3 -- Identifying inbound data (Organizational)
    \item I4 -- Identifying context (Organizational/External)
    \item I5 -- Identifying shared data (Receiving party)
    \item I6 -- Identifying stored data (Organizational)
    \item I7 -- Identifying retrieved data (Organizational, [Future] receiving party)
\end{itemize}
\end{tcolorbox}

Closely related, as a special case of linkability, are \textbf{IDENTIFIABILITY threats} including another risk that we consider critical, I6 IDENTIFYING STORED DATA. \acrshort{bci}-acquired data being stored may be identified (linked to the user) because they are insufficiently minimized/anonymized before storage; analogously I7 IDENTIFYING RETRIEVED DATA applies to data passed on to third parties. These threats were considered critical as they represent the basic risk of data leakage and theft, i.e., that brain-signal-derived personal information can be linked back to an individual. I1 IDENTIFYING CREDENTIALS, I2 ACTIONS IDENTIFY USER, and I3 IDENTIFYING INBOUND DATA we considered as medium risk threats.

\begin{tcolorbox}[colback=TealBlue!5!white,colframe=TealBlue!75!white]
\textbf{REPUDIATION threats}: 
\begin{itemize}
    \item Nr1 -- Credentials non-repudiation (External)
    \item Nr2 -- Non-repudiation of sending (Organizational)
    \item Nr3 -- Non-repudiation of receipt (Organizational)
    \item Nr4 -- Non-reputable storage (Organizational)
    \item Nr5 -- Non-repudiation of retrieved data ([Future] receiving party)
\end{itemize}
\end{tcolorbox}

The next category, \textbf{NON-REPUDIATION threats}, is especially interesting in the \acrshort{bci} context. As mentioned above, in contrast to the security context, non-repudiation is considered a threat for privacy. From a privacy perspective, non-repudiation ensures that one can, for example, plausibly deny having logged into a questionable website. However, as stated in the \acrshort{gdpr}: \emph{\say{The right to the protection of personal data is not an absolute right; it must be considered in relation to its function in society and be balanced against other fundamental rights, in accordance with the principle of proportionality}} (Recital 4 \acrshort{gdpr}). From a security perspective, non-repudiation ensures that users cannot arbitrarily shift the blame for their actions to a \acrshort{bci} - \emph{\say{I did not do that - this was my \acrshort{bci} misinterpreting my intentions!}}). Weighing the privacy and security considerations, we are convinced that here security prevails, thus, we have not assigned any risks in this threat category (see also~\nameref{sec:userautonomy} section).  

\begin{tcolorbox}[colback=TealBlue!5!white,colframe=TealBlue!75!white]
\textbf{DETECTABILITY threats}: 
\begin{itemize}
    \item D1 -- Detectable credentials (External)
    \item D2 -- Detectable communication (External)
    \item D3 -- Detectable outliers (External)
    \item D4 -- Detectable at storage (External [with access to the system])
    \item D5 -- Detectable at retrieval ([Future] receiving party)
\end{itemize}
\end{tcolorbox}

The next category are \textbf{DETECTABILITY threats}, which refer to the ability to detect whether an item of interest exists or not, without having access to the data. An example given in LINDDUN-Go: \emph{\say{By detecting that a celebrity has a health record in a rehab facility, one can infer the celebrity has an addiction, even without having access to the actual record.}} In our view, this threat category is neither specific nor critical for current \acrshort{bci} applications and we have not assigned any substantial risk to the five threats in this category.  

\begin{tcolorbox}[colback=TealBlue!5!white,colframe=TealBlue!75!white]
\textbf{UNAWARENESS threats}: 
\begin{itemize}
    \item U1 -- No transparency (Organizational)
    \item U2 -- No user-friendly privacy control (Organizational)
    \item U3 -- No access or portability (Organizational)
    \item U4 -- No erasure or rectification (Organizational)
    \item U5 -- Insufficient consent support (Organizational)
\end{itemize}
\end{tcolorbox}

This is in contrast to \textbf{UNAWARENESS threats}, namely U1 NO TRANSPARENCY, U2 NO USER-FRIENDLY PRIVACY CONTROL (and related also Nc4 AUTOMATED DECISION MAKING, see below). We argue that unawareness may be the most dangerous source of risks in the present stage of \acrshort{bci} technology; in our view it is fundamentally impossible to exert the right to informational self-determination in a meaningful way if brain data of unknown information content are processed by opaque algorithms implemented in closed-source proprietary software - a black box within a black box within a black box. We advocate to \textbf{"Avoid security through obscurity"}, relying on transparent and interpretable solutions (cf. Tesla’s open-source patent philosophy for autonomous car security software~\cite{wang_study_2020}). U3 NO ACCESS OR PORTABILITY and U4 NO ERASURE OR RECTIFICATION we consider as additional high risk unawareness-related threats. 

\begin{tcolorbox}[colback=TealBlue!5!white,colframe=TealBlue!75!white]
\textbf{NON-COMPLIANCE threats}: 
\begin{itemize}
    \item Nc1 -- Disproportionate collection (Organizational)
    \item Nc2 -- Unlawful processing (Organizational)
    \item Nc3 -- Disproportionate processing (Organizational)
    \item Nc4 -- Automated decision making (Organizational)
    \item Nc5 -- Disproportionate storage (Organizational)
\end{itemize}
\end{tcolorbox}

The final threat category addressed by the LINDDUN framework is \textbf{NON-COMPLIANCE threats} which arise if the system does not comply with data protection principles. The \acrshort{gdpr}, for example, establishes 6 data protection principles including purpose limitation, data minimization, integrity and confidentiality (as well as accountability as an overarching seventh principle related to the other six). In the LINDDUN framework, this category has been developed mainly in the context of the EU’s \acrshort{gdpr}, but the underlying principles may be considered independent of a specific geographic or legal domain. On the other hand, threat manifestation in this category appears highly context-dependent, for example, contingent on differences in legal and political factors. Assessments of non-compliance-related threats in \acrshort{bci} applications, due to their highly context-dependent nature, can only be preliminary and in our view indicate a threat potential rather than an actual threat. Specific threats Nc1-5 are DISPROPORTIONATE COLLECTION, UNLAWFUL PROCESSING, DISPROPORTIONATE PROCESSING, AUTOMATED DECISION MAKING, and DISPROPORTIONATE STORAGE. Except for Nc2 (unlawful processing) we consider all non-compliance threats as \acrshort{bci}-specific (due to the specific nature of the collected and processed data) and have assigned a high risk level (in the sense of a general risk potential rather than an actual risk, see above). Nc2 Unlawful processing we consider as non-\acrshort{bci}-specific as long there are no specific  \verb|"BCI laws"| in place (but cf. the introduction of \verb|"NeuroRights"| in Chile~\cite{lorena_guzman_h_chile_2022}). 

In addition, we also consider \textbf{DISCLOSURE OF INFORMATION threats} which imply risks to brain data confidentiality, as (brain) privacy highly depends on this core aspect of security. Associated risks may exist across the whole \acrshort{bci} data flow, including all stages of data at rest and data in transit. Measures to ensure brain data confidentiality will be discussed in the~\nameref{sec:hide} Strategy Section.

Fig.~\ref{fig:dfd} shows a summary of the \acrshort{bci}-specific privacy threats as discussed above mapped to their hotspots in the \acrshort{bci} flow diagram, including \textbf{several \acrshort{bci}-specific, critical risk-level threats.}

\begin{figure}[h!]
\begin{centering}
    {\includegraphics[scale=0.38]{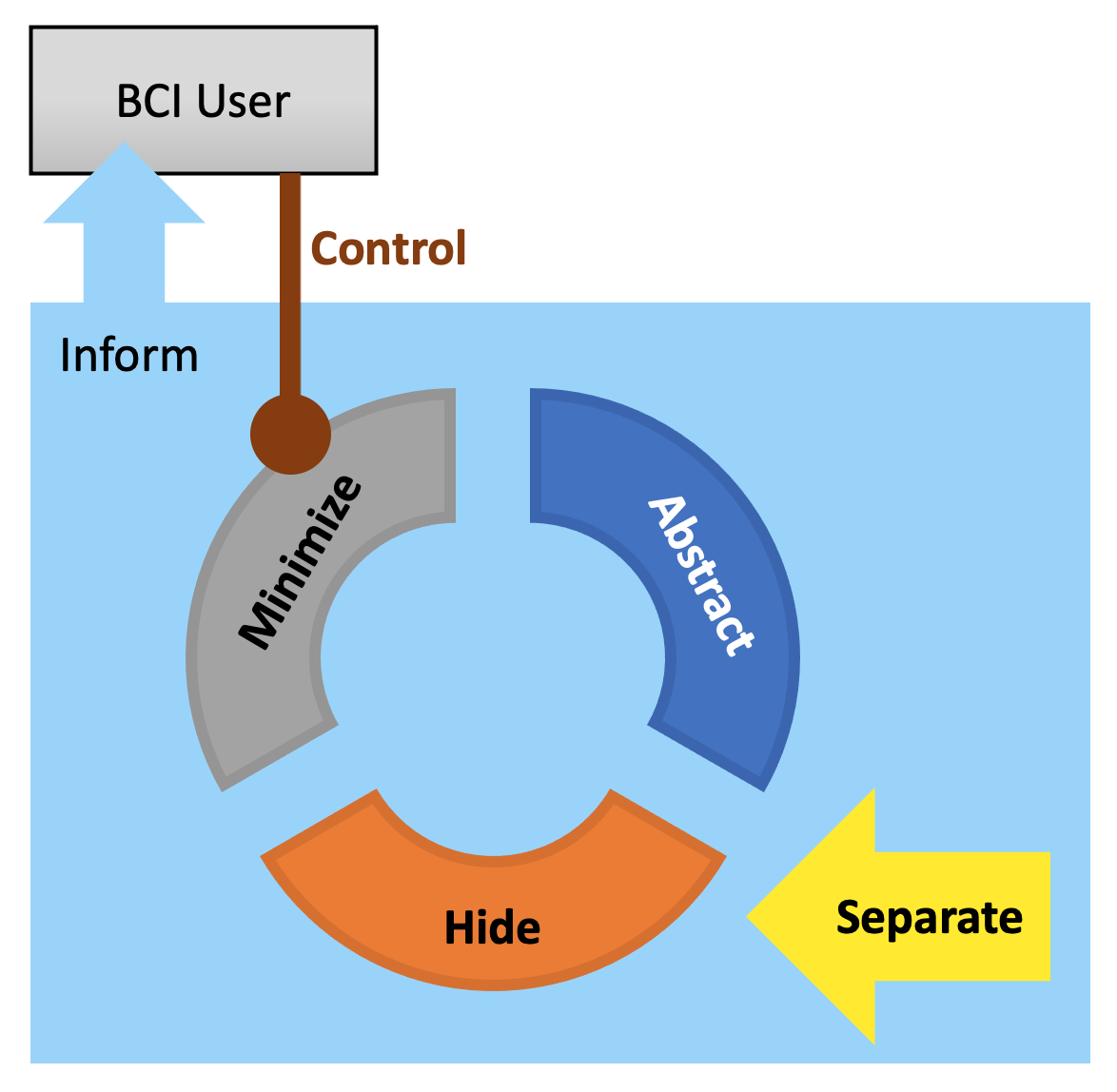}}
    \caption[Privacy design patterns]{\textbf{Privacy design patterns}  after Hoepman~\cite{blue-book}, in our color-coding (also used in the following figure). Note that the circular arrangement here was chosen to symbolize that these strategies “protect” the processing nodes (see Fig.~\ref{fig:mapping-hoepman}). }%
    \label{fig:design-patterns}
\end{centering}
\end{figure}

In the next step, to develop suitable mitigation strategies addressing this threat landscape, we need strategies that can be used modularly for flexible application scenarios, and which lead to actionable tactics, including descriptions, requirements, and implementation aspects regarding the different levels to be considered, such as suitable \acrfull{pet}. 

\section{Threat mitigation by privacy engineering}

We selected the privacy engineering approach by Hoepman~\cite{hoepman-privacy_2014}. This methodology has the advantage that it covers the whole range of \acrshort{gdpr} privacy requirements: In the review by (Huth \& Matthes 2019~\cite{huth-gdpr}), the Hoepman privacy engineering approach was the only one with \emph{\say{\acrshort{gdpr} completeness}}. In contrast to the also recent PRIPARE methodology by Notario et al.~\cite{notario_pripare_2015}, Hoepman also offers explicit links to specific techniques for implementation.

\begin{figure}[h!]
\begin{centering}
    {\includegraphics[scale=0.4]{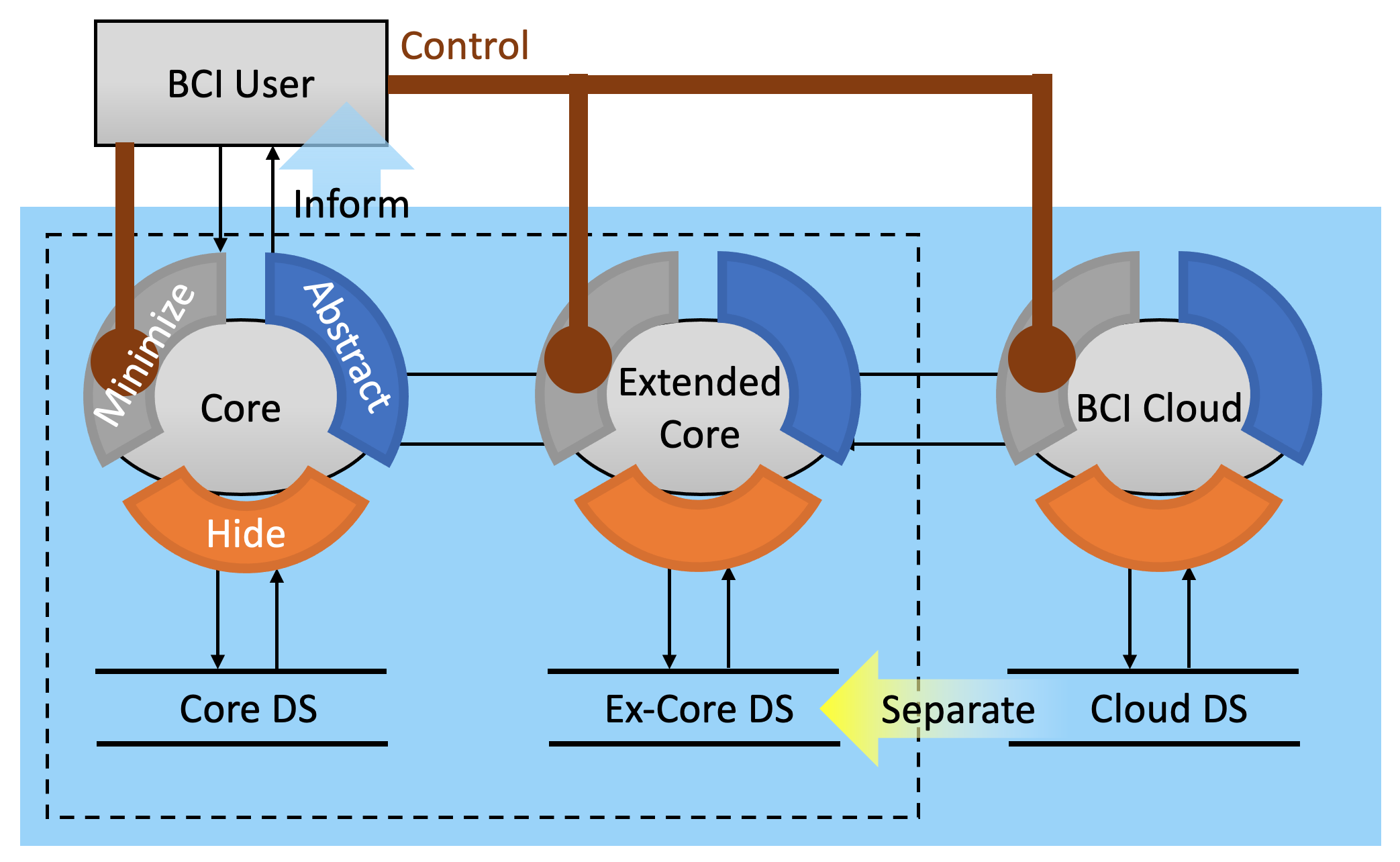}}
    \caption[Mapping of privacy engineering strategies onto the BCI DFD.]{\textbf{Basic neuroprivacy and -security
architecture} mapping the Hoepman privacy engineering strategies onto the BCI data flow. Minimize, abstract, and hide strategies may apply to processing in core, extended core, and cloud components. The Inform strategy plays a fundamental role in our architecture and pervades all levels and components of the BCI system, corresponding to the critical risk of unawareness threats. The main role of the Separate strategy lies in the transition of centralized to decentralized solutions (see main text for further details). }%
    \label{fig:mapping-hoepman}
\end{centering}
\end{figure}

The Hoepman methodology is based on 8 privacy design strategies which are derived from the \acrfull{oecd} privacy guidelines, the draft of the \acrshort{gdpr}, as well as the ISO 29100 privacy framework (Fig.~\ref{fig:design-patterns}). Specifically, the design strategies comprise 4 more technical, \textbf{data processing-oriented} strategies: Minimize, Hide, Separate, and Abstract; as well as 4 \textbf{process-oriented} strategies: Inform, Control, Enforce, and Demonstrate, which deal with the organizational and regulatory aspects of privacy engineering. Crucially, each strategy comes with a set of \textbf{tactics} guiding the concrete implementation of the various strategies. Fig.~\ref{fig:mapping-hoepman} shows a mapping of these strategies onto our \acrshort{bci} flow diagram (note that the strategies Demonstrate and Enforce, which are directed to the data controller, will be discussed in the context of legal and regulatory aspects). 

The abstract flow diagram combined with the mapped engineering strategies provides \textbf{general architecture for \acrshort{bci}s to protect brain privacy}. This architecture can be applied to concrete \acrshort{bci} examples in a flexible way, can be extended for future use cases, and offers a systematic link to concrete implementation tactics, as we discuss in the following for the individual design strategies. 

\bigskip

%% file: figures/7-architecture/fig06-flowchart.tex
\begin{figure}[t]
\begin{centering}
    {\includegraphics[scale=0.4]{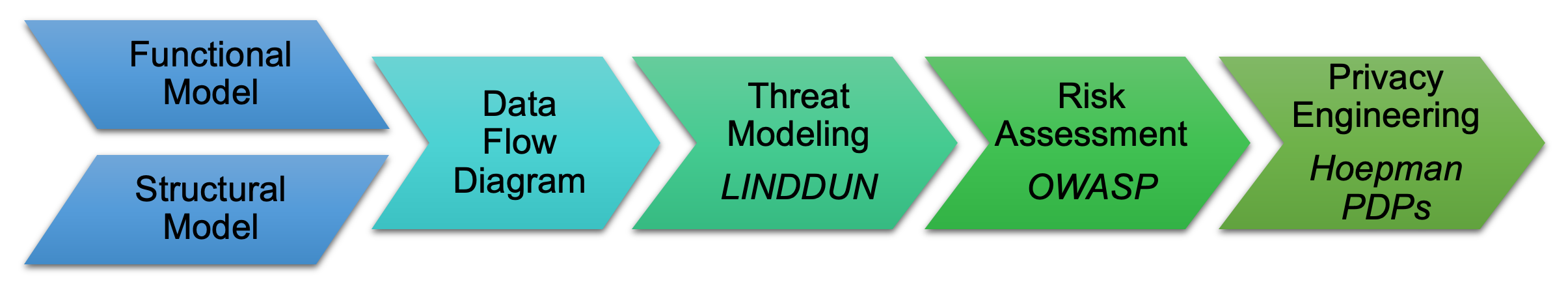}}
    \caption[Flowchart of analysis  steps and related methodologies.]{\textbf{Flowchart of analysis  steps and related methodologies.} }%
    \label{fig:flow-chart}
\end{centering}
\end{figure}

%% file: figures/7-architecture/fig07-dfd.tex
\begin{figure}[t]
\begin{centering}
    {\includegraphics[scale=0.4]{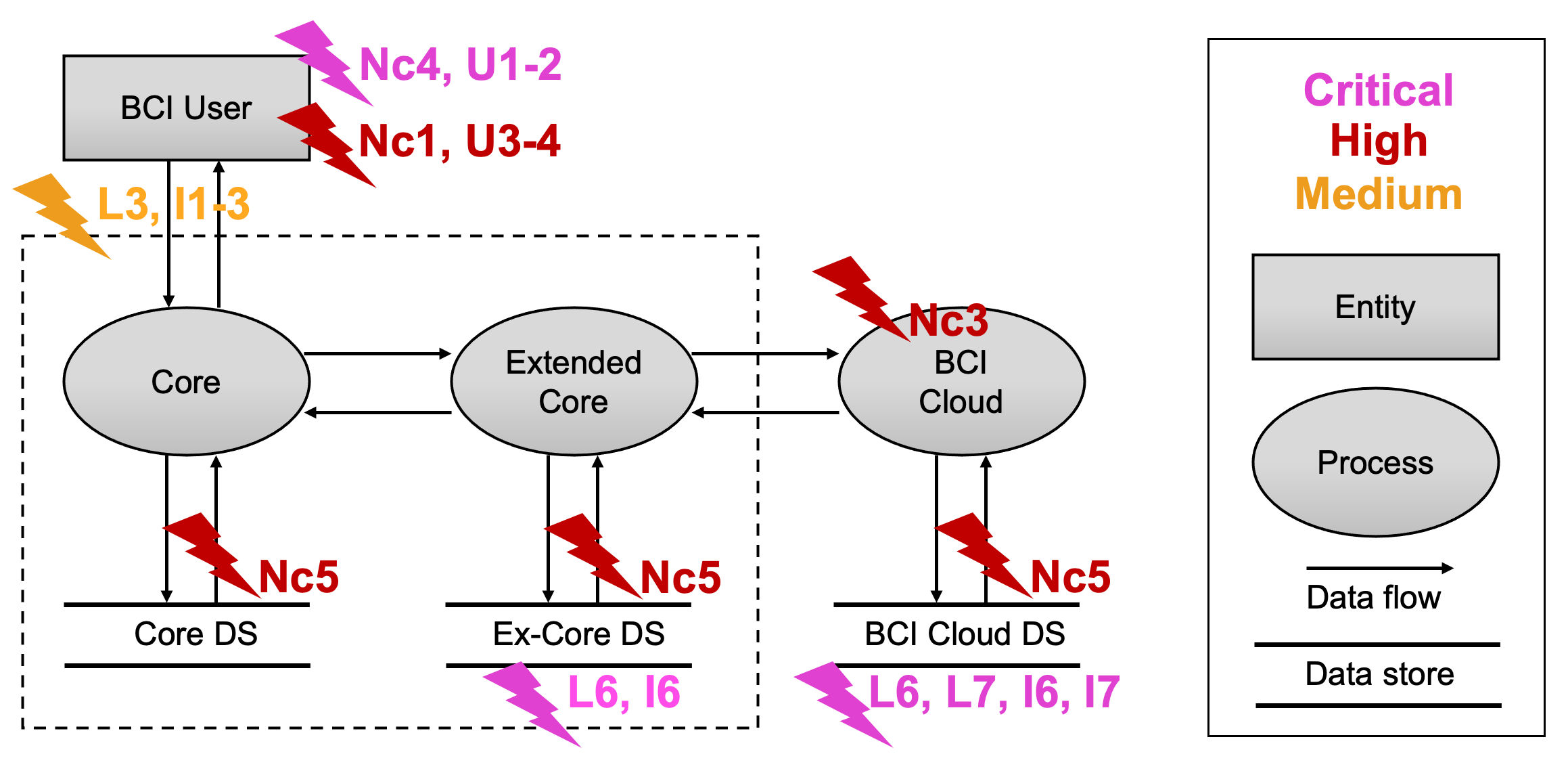}}
    \caption[BCI data flow diagram (DFD)]{\textbf{BCI data flow diagram (DFD) and BCI-specific privacy threats mapped to their potential hotspots}  (here we assume global functions to run in the cloud). Legend of risk levels and \acrshort{dfd} elements on the right. Dashed line indicates one possible trust boundary. Threat categories; L: Linkability; I: Identifiability; U: Unawareness; Nc: Noncompliance. Detailed threat acronyms: See main text.}%
    \label{fig:dfd}
\end{centering}
\end{figure}

%% file: chapters/6-privacy-designs-strategies.tex
\chapter{Privacy Design Strategies for BCIs}\label{chap:design-strategies}
In this chapter, we will discuss how the 8 privacy design strategies as introduced above can be applied to \acrshort{bci} applications. We will highlight both actionable design features as well as research questions: 
\begin{tcolorbox}[colback=green!5!white,colframe=green!75!black]
\textbf{Actionable design features}: After the discussion of the strategies we include boxes with concrete privacy-by-design \acrshort{bci} components which do not require extensive further research to be already useful for \acrshort{bci} privacy-preservation. 
\end{tcolorbox}

\begin{tcolorbox}[colback=magenta!5!white,colframe=magenta!75!black]
\textbf{Research questions}: We also highlight selected research questions related to privacy-preserving \acrshort{bci} techniques and methods which may be helpful or required for the transformation of today’s emerging \acrshort{bci} technologies to a mature and productive technology stage (c.f., Fig.~\ref{fig:bci-trls}). 
\end{tcolorbox}

\section{Minimize}
The Minimize strategy requires limiting as much as possible the collection and the processing of personal data: \emph{\say{The most obvious strategy to protect privacy is to minimise the collection of personal data. Nothing can go wrong with data you do not collect.}}~\cite{blue-book}. The Minimize strategy has the implementation tactics \emph{Select, Exclude, Strip} and \emph{Destroy}. 

For \acrshort{bci} applications, the most basic and straightforward implementation of the Minimize strategy would be to discard the recorded neurodata after they have been passed through the decoder - during intended use, the neurodata do not leave the core \acrshort{bci} cycle. It has been argued that under such conditions, a \acrshort{bci} used for cursor control would raise few, if any, privacy issues as it would not reveal any more personal information than occurs when controlling a cursor with an ordinary computer mouse\footnote[9]{However, even in this case privacy issues may arise, e.g., through unsecure wireless data transmission~\cite{karim_trusted_2019}}~\cite{ibm-paper}. 

However, such minimal data storage may have disadvantages for the development and operation of safe and performant \acrshort{bci} applications. It would prohibit large-scale data pooling, which however may be required to develop improved decoding models at least in the current stage of \acrshort{bci} technology, for example in the context of exploratory algorithm development, research on robust subject-independent models or on meta-decoders. Also certain variants of important techniques such as adaptive decoding~\cite{chen2018on, batch-size} and data-driven anomaly detection may require storage of data, although not necessarily permanent, but at least over shorter time horizons. Thus, another concept for implementing the \emph{Select} tactics is storing the data, but applying a \textbf{feature limitation} approach to the stored data (also following the EU data minimization principle). In this approach, only a limited set of features of the neurodata is stored, or certain features are removed from the data before storing.

Although this seems an attractive idea, the feature limitation in the context of \acrshort{eeg} data and \acrshort{bci}s is not necessarily straightforward and may require further research. Questions here would for example be \emph{\say{Is it possible to do reliable feature limitations (and if yes, to what extent) for brain signals?}}, \emph{\say{Which information is relevant and irrelevant for the task?}}. As an example, one can think of limitations on the sensor level (recording only a subset of electrodes relevant to the task). For example, the NextMind \acrshort{bci} is based on visually-evoked potentials and fitting to this signal source, all electrodes are positioned above the occipital area containing the primary and secondary visual cortex. However, \acrshort{eeg} electrodes positioned in such a way, e.g., above the visual or motor areas, may also pick up signals from brain areas involved in cognitive processes, e.g., speech processing.

Therefore, a next logical step is to take the feature limitation approach from sensor space to the signal space, and to strip signal components, such as by filtering out certain frequency ranges, etc. Feature selection is the process of selecting an optimized subset of features for \acrshort{ml} model training and a broad spectrum of different algorithms has been developed for this purpose~\cite{james_introduction_2013}. As a \acrshort{bci} example, filter banks are an important component of many classical pipelines and can also be integrated, e.g., with deep networks for \acrshort{eeg}~\cite{mane_fbcnet_2021}.

The concept of the so-called \acrshort{bci} Anonymizer\footnote[10]{Note that the concept of the BCI anonymizer refers to online BCI usage, in contrast to EEG anonymization for offline analysis, see Section 9. }~\cite{bonaci_phd_2015, bci-anonymizer} is based on such a feature limitation approach. This tool has been proposed as a way to remove private information from \acrshort{eeg} data before they are stored or transmitted. The hypothesis here is that recorded brain signals can be decomposed into a collection of characteristic components in real-time. From these components, one wishes to extract exactly the information needed for the \acrshort{bci} application while filtering private information out. Several issues with the \acrshort{bci} Anonymizer idea have been identified, including resource constraints in \acrshort{bci} devices, lack of access to proprietary algorithms, lack of a clear method for separating private information from intentions and a general lack of any implementation details. Despite the attractiveness of the \acrshort{bci}-anonymizer idea, appropriate methods for selectively removing all sensitive information from \acrshort{eeg} data are not yet available, and it appears unlikely that this can be achieved by simple removal of (additive) components, reflected by the abandoned status of the \acrshort{bci} Anonymizer patent application~\cite{bci-anonymizer}.

In contrast, a promising yet methodologically more complex approach to feature limitation leverages \textbf{generative \acrshort{ml}} methods. Yao and colleagues have adopted  domain adversarial \acrshort{dl} techniques for feature removal from \acrshort{eeg}. Generative \acrshort{ml} methods such as \acrlong{gan}s (\acrshort{gan}s) can solve a wide range of tasks. For example, they can generate photorealistic images from high-dimensional noise~\cite{karras_progressive_2017}, generate segmentation masks for medical imaging data~\cite{navab_u-net_2015}, translate photographs into an image with a learned artistic style~\cite{Gatys_2016_CVPR}, or generate arbitrary images from verbal instructions~\cite{dalle2}. Similarly, generative methods can be set up to translate a signal into a corresponding one of the same dimensions, but enforcing that information such as age or gender cannot be inferred from the generated one. For example, Yao et al. report that using a model structure with multiple generators and discriminators it was possible to simultaneously achieve state-of-the-art accuracy on a public \acrshort{eeg} data set, while removing privacy-related features~\cite{yao_improved_2019}. 

In summary, the Minimize strategy offers a wide range of tactics ranging from simple and immediately applicable to highly complex, and should always be evaluated for a \acrshort{bci} application, as early as in the current \verb|"Emerging Technology Phase"|. Feature limitation techniques based on generative \acrshort{ml} are a highly promising research priority for the coming transition phase to more mature \acrshort{bci} solutions.

\begin{tcolorbox}[colback=green!5!white,colframe=green!75!black]
\textbf{BCI-Limiter}: Integrating channel pre-selection and feature selection, such as using optimized filter banks, for feature limitation. 
\end{tcolorbox}

\begin{tcolorbox}[colback=magenta!5!white,colframe=magenta!75!black]
How can \textbf{generative \acrshort{ml}} be used, likely combined with \textbf{multi-objective optimization} (to balance extent of feature reduction with computation and performance costs), to achieve fine-grained feature limitation with minimal performance and computation costs and intuitive control not only by domain experts, but also \acrshort{bci} users?  
\end{tcolorbox}

\section{Abstract}

The Abstract strategy aims at limiting as much as possible the detail in which personal data are processed. \emph{\say{The less detailed a personal data item is, the more we ‘zoom out’, the lower the privacy risk is.}}~\cite{blue-book}. The Abstract strategy includes the tactics \emph{Summarise}, \emph{Group}, and \emph{Perturb}, for processing personal data in a more coarse-grained or aggregate manner. Examples are replacing the exact birth date by an age category, e.g., rounded to years, or a full address by the city of residence. 

Application of the \textbf{Abstract strategies to the meta-data} acquired and stored alongside the brain data is straightforward and natural, and may provide substantial privacy advantages. These steps are directly applicable when meta-data are stored alongside the brain signals. For example, if handedness is determined with a handedness questionnaire such as the widely-used modified Oldfield questionnaire~\cite{oldfield_assessment_1971}, just the resulting handedness index might be stored, not the answers to all questions. Another example would be subject age, where exact dates of birth could be replaced by age categories, for example in applications involving age as a covariate or aiming at \verb|"brain age"| decoding~\cite{engemann_reusable_2021}.

Applications of the \textbf{Abstract strategies to the brain data} themselves appear less straightforward, as there is a conceptual closeness to other strategies such as feature limitation which is considered part of the minimization strategy (see above). The Abstract strategy might seem naturally implemented in feature-based brain signal decoding pipelines. For example, Riemannian-Geometry-based decoders (among the currently most successful class of decoders for \acrshort{eeg} \acrshort{bci}s) use the covariance matrix of the input signals as their input, thus the input is not a time series anymore, \verb|"abstracting"| the detailed temporal structure in the covariance structure of the data. However, the same approach could also be seen as an example of feature limitation (all features except the covariance are omitted). Another example is the so-called \acrfull{lfc} which, in particular combined with an \acrshort{rlda} decoder for intracranial \acrshort{eeg} data, can be highly informative, e.g., about hand movement kinematics~\cite{hammer_role_2013}. Again, the \acrshort{lfc} could be seen as an abstraction in the sense of \verb|"zooming out"| (reducing  detail due to the low-pass filtering of the data), or as a feature limitation (omitting the high-frequency features of the brain signals). For brain signals, Abstract strategies might be best seen as a special case of limitation, namely \textbf{coarse-graining feature limitation.}

Closer to the concept of a privacy-preserving abstraction, and an interesting topic for research in the domain of brain signals, could be learned privacy-preserving \textbf{low-dimensional embeddings} of the input data; for example Bezzam and colleagues~\cite{bezzam_learning_2022} demonstrated for lensless cameras the privacy-preserving potential of joint optimization of an image classifier together with rich (allowing accurate classification) but lower-resolution optical embeddings learned already at the sensor level of the camera. On the other hand, such techniques - adapted to the brain signal domain - could also be included in the Hide strategy category together with other kinds of \acrshort{ppml} (see following section). 

In summary, privacy engineering using the Abstract strategy is straightforward and important with respect to the meta-data acquired and stored alongside the brain data; at least current brain data processing techniques which possess an abstraction aspect seem to more naturally fit into related categories (Minimize, Hide).   

\begin{tcolorbox}[colback=green!5!white,colframe=green!75!black, enlarge top by=1cm, enlarge bottom by=1cm]
\textbf{BCI-MetaAbstract}: Use suitable abstractions for \acrshort{bci} metadata when possible. 
\end{tcolorbox}

\begin{tcolorbox}[colback=magenta!5!white,colframe=magenta!75!black]
Closely related to the research question on limitation methods - How can \acrshort{ml} approaches, combined with advanced optimization, be used for performance- and computation-efficient abstraction not only on meta- and behavioral data, but also for \textbf{abstraction on the brain or multimodal signal level}? It may not be optimal to implement limitation and abstraction methods separately, but rather jointly, allowing joint optimization.
\end{tcolorbox}

\section{Hide}\label{sec:hide}

This important strategy comprises a broad spectrum of privacy-enhancing tactics and techniques employed to ensure the \textbf{confidentiality}, \textbf{unlinkability}, and \textbf{unobservability} of personal data. Its logical place is downstream to the Minimize and Abstract strategies as discussed above: After it has been decided that personal data need to be recorded, and at which level of abstraction, the aim of the Hide strategy is to make sure that these data do not become public or known. 

Importantly, this strategy also addresses confidentiality, which can be seen as a security rather than a privacy topic (see footnote[8] Section~\ref{sec:privacy_threat}), but due to its pivotal role for privacy we consider it crucial to include this aspect in a brain-privacy-centered framework as well. Tactics within the Hide strategy are to \emph{Restrict}, \emph{Obfuscate}, \emph{Dissociate}, and to \emph{Mix}. From the implementation side, access control policies, encryption, anonymization (removing directly-identifying data), and a large part of the rapidly-growing \acrshort{ppml} field all fall into this broad category. In the following we discuss key aspects for the adoption of the Hide strategy to \acrshort{bci} applications, focusing on (i) brain data anonymization and its relation to \acrshort{eeg} biometrics, (ii) \acrshort{ppml} for \acrshort{bci}s, and (iii) the crucial role of unlinking brain data and contextual data. 

\subsection{Brain data: Anonymization and re-identification}

Can brain data be anonymized, and if yes, how? This is a topic of paramount importance, as Recital 26 EU \acrshort{gdpr} states that the principles of data protection should apply only to information concerning an identified or identifiable natural person (personal data), and not to anonymous information (non-personal data). If brain data can be anonymized, this could greatly facilitate large-scale data pooling, for example. To determine whether a natural person is identifiable, \acrshort{gdpr} refers to the \textbf{risk of identification}, considering all means which are \emph{\say{reasonably likely}} given factors such as costs, time, current and prospective technology. What does this mean and what are the consequences for brain data recorded by \acrshort{bci} systems? 

Finck and Pallas~\cite{finck_they_2020} have recently examined the concept of personal vs. non-personal data under the \acrshort{gdpr} from both a legal and computer science perspective. They point out that there is a basic contradiction between the \acrshort{gdpr} on the one hand, which implies an acceptable residual risk of identification compatible with the anonymous status of data, with interpretations by national supervisory authorities as well as the European Data Protection Board on the other, which consider that no remaining risk of identification is acceptable for data to qualify as anonymous. Purtova has provided a similar analysis of the uncertainties surrounding the concept of identification and identifiability as critical boundary concepts of data protection law~\cite{purtova_knowing_2022}. Another open thread of legal discussion concerns the question whether the scope of personal data protection should be extended to \acrshort{ml} models, such as \acrshort{dl} networks trained on personal data~\cite{leiser_governing_2020}.  

On the technical side, what can we say about the feasibility and risk of (re-)identification of brain data, \acrshort{eeg} data in particular? Here, the literature on \acrshort{eeg} for biometrics is helpful, as this literature provides insights into the feasibility of re-identification of an otherwise anonymous \acrshort{eeg} recording against a database of known \acrshort{eeg}s\footnote[11]{However, see Bigdoly et al.~\cite{bidgoly_towards_2022}, for an EEG-based authentication approach based on a privacy-preserving EEG fingerprint function.}. The \acrshort{eeg} data used for biometrics may be spontaneous ongoing activity, or may also contain different kinds of event-related responses, such as to \verb|"passthoughts"|. As a large number of studies have reported high biometric accuracies, the feasibility of \acrshort{eeg} re-identification is sometimes taken as a fact. However, how well established is it that \acrshort{eeg} recordings can be re-identified on a large scale of individuals, across large time scales, and with high accuracies? 

According to the survey by Bigdoly and colleagues, only 13\% of the included studies used \acrshort{eeg} from as many as 100 to 150 participants, while arguably an \acrshort{eeg} biometric system would have to be functional in a population several orders of magnitude larger in size. 27\% of the reviewed studies used 10 subjects or even fewer~\cite{bidgoly_towards_2022}. As also stated by Bigdoly and colleagues, it has been argued that the accuracy of \acrshort{eeg} biometrics may decrease as a function of an increasing number of participants (c.f. also~\cite{hutchison_personal_2010}). Also the number of studies investigating the long-term stability of \acrshort{eeg} biometrics is relatively small. One notable exception is for example the study by Ruiz-Blondet et al., who re-tested their “Cognitive Event RElated Biometric REcognition” (CEREBRE) protocol in 20 subjects after 48 to 516 days, reporting 100\% recognition accuracy~\cite{ruiz-blondet_permanence_2017}. Other factors which may challenge the real-world performance of \acrshort{eeg} biometrics are changes in affective, vigilance, and hormonal state, medication, sobriety, occurrence (or disappearance) of neurological disorders, all of which can change the recorded \acrshort{eeg} patterns. Thus, it may be argued that despite the impressive reports in the \acrshort{eeg} biometrics literature and encouraging results, at present it is unclear what accuracy re-identification/biometrics of \acrshort{eeg} will be able to achieve on large scales (10s of thousands of users or more, over years). It will be crucial to find out whether these accuracies at scale will be high enough to enable reliable \acrshort{eeg}-based identification and authentication, or, on the other hand, low enough to dispel privacy concerns due to the risk of \acrshort{eeg} re-identification.

Furthermore, even if the raw \acrshort{eeg} recordings can be re-identified, feature limitation approaches may be employed to reduce or remove personally-identifying \acrshort{eeg} features. As discussed above, the concept of the so-called \acrshort{bci} Anonymizer was based on a simple feature limitation approach. Generative \acrshort{ml} (as also discussed above with respect to feature limitation in general), however, may help to better remove personally-identifying \acrshort{eeg} components. \textbf{Large scale studies} with substantially more subjects will be needed for the development and evaluation of advanced \acrshort{eeg} anonymization algorithms, assessment of the risk of identification, and, in parallel with advances on the legal side, clarifying the personal vs. non-personal status of data processes with different anonymization approaches. 

\subsection{Privacy-Preserving Machine Learning (PPML) for BCI applications}

\acrshort{ppml} is a class of \acrshort{ml} specifically designed to mitigate data privacy concerns\footnote[12]{Many of the strategies adopted in \acrshort{ppml} fit to the Hoepman “Hide strategy”, therefore we discuss PPML for EEG and BCI applications as a whole in this section. However, PPML is a wide field comprising many different techniques across the whole ML pipeline and life cycle. Addition of noise is systematically exploited by differential privacy PPML, which would fit to the perturb tactics in the Hoepman Abstract Strategy. Hoepman also discusses homomorphic encryption in the Abstract Strategy section, and SMC in the Separate Strategy section, respectively~\cite{blue-book}. We would argue that the assignment of methods to strategies can be ambiguous and should not be overcomplicated; what we want to encourage is that the full range of advantages and solutions offered by state-of-the-art PPML be evaluated and included in a systematic BCI privacy engineering strategy.   }~\cite{xu_privacy-preserving_2021}. This field for example includes research on methods for \acrshort{dl} on encrypted data (e.g., CryptoNets~\cite{pmlr-v48-gilad-bachrach16}). Recently, studies have started to explore the potential of \acrshort{ppml} for \acrshort{eeg} data analytics. For example, Popescu et al. have used \textbf{homomorphic encryption}, designed for model inference on encrypted data and giving encrypted predictions decipherable only by the data owner, for privacy-preserving classification of \acrshort{eeg} data. They report comparable prediction performance of the models operating on the encrypted and plaintext data, but increased computational time for training in the homomorphic encryption case~\cite{popescu_privacy_2021}. 

Agarwal et al. proposed cryptographic protocols based on \textbf{\acrfull{smc}} for \acrshort{eeg}~\cite{agarwal_protecting_2019}. \acrshort{smc} is a specialty of cryptography that studies methods to allow a party of participants joint computation of a public function while keeping their respective inputs private. In the cited work, Agarwal et al. utilized \acrshort{smc} for estimating drowsiness of drivers using linear regression over \acrshort{eeg} signals from multiple users in a fully privacy-preserving manner and report a manageable computational cost. 

A series of recent works has studied the feasibility and performance of \textbf{federated learning} for \acrshort{eeg}~\cite{gao_hhhfl_2020, ju_federated_2020, Szegedi2019EvolutionaryFL, sun_scalable_2022}, such as for online seizure detection in epilepsy~\cite{baghersalimi_personalized_2022}. Federated learning addresses similar privacy scenarios as \acrshort{smc}, but instead of cryptography relies on \acrshort{ml} techniques to train \acrshort{ml} models, such as deep neural networks, on distributed data, by training local models on the local, private data, and providing methods to combine the local models into a global model without the need to disclose the local data. 

Others have recently started to explore \acrshort{ppml} for brain data using \textbf{synthetic data generation}. For example, \acrshort{gan}s are a novel, powerful generative \acrshort{ml} model class that can be used to generate synthetic yet realistic \acrshort{eeg} data~\cite{hartmann_eeg-gan_2018}. Debie and Mustafa have proposed a \acrshort{gan} trained under a differential privacy model for generating privacy-preserving synthetic \acrshort{eeg} data~\cite{debie_privacy-preserving_2020}. Similarly, Pascual and colleagues proposed a \acrshort{gan} model to generate synthetic seizure-like \acrshort{eeg} signals used to train seizure detection algorithms, reporting that their \acrshort{gan} model was both successfully privacy-preserving and without performance degradation during seizure monitoring~\cite{pascual_epilepsygan_2021}.

In summary, there are first encouraging steps highlighting the potential of \acrshort{ppml} in the brain signal domain. The application of \acrshort{ppml} in this context is a very recent development -- all of the works summarized above were published not earlier than 2019. It is important to underscore the fact that \acrshort{ppml} for brain signals is still a research-level technology in a very early stage. There still seems to be a long way to go to implement these approaches in commercial products. \acrshort{ppml} in itself is also a highly dynamic field of research with new, promising solutions being generated which have not yet been investigated in the \acrshort{bci} domain. For example, Mo et al. have recently proposed privacy-preserving federated learning with \acrfull{tte}, showing the feasibility of running deep networks in \acrshort{tte} on the client side for local training, and on the server side for secure aggregation, thus strongly hiding model updates from adversaries~\cite{mo_ppfl_2021}. 

\begin{tcolorbox}[colback=magenta!5!white,colframe=magenta!75!black]
Given the convergence of two dynamic technologies, neurotechnology/\acrshort{bci} and \acrshort{ppml}, in our view, at the present stage of development it would be by far premature to prescribe any concrete \acrshort{ppml} technique as a general solution to \acrshort{bci} privacy concerns. Rather, we think that \textbf{research on a broad spectrum of \acrshort{ppml} solutions for \acrshort{bci} applications will be a key component in the transformation of the \acrshort{bci}} field from its current, early phase to a more mature level in the future (Fig.~\ref{fig:bci-trls}). Such research will enable informed choices which \acrshort{ppml} techniques are practicable applicable for online \acrshort{bci}s considering their potential computation and performance costs, and relatedly, which \acrshort{ppml} techniques may or may not be necessary and/or useful for specific \acrshort{bci} applications. 
\end{tcolorbox}

\subsection{Unlinking brain data and contextual data}
Unlinkability is one of the main concepts within the Hide Strategy. In this last section within the Hide Strategy chapter, we want to draw attention to the important role of unlinkability in the \acrshort{bci}/brain privacy context. This is due to the fact that in most cases, decoding information from brain signals also requires synchronized information about the sensory or behavioral context. For example, trying to decode the emotional reaction to different faces requires knowledge about which face the subject was looking at when. Trying to steal PIN codes by presenting numbers also requires knowledge about time and identity of the numbers presented. Brain data without context contain much less decodable personal information (still decodable may be information about (approximate) age, gender, or neurological disorders). Thus unlinking brain and contextual data provides a powerful means to substantially reduce the risk of information leakage. 

In contrast to the complex and methodologically demanding techniques discussed above, such unlinking could be implemented with much simpler means and already \verb|"here and now"|, for example by systematically storing brain data and context data in separate files, without mutual identifiers, and keeping the linking information encrypted. We believe that this approach is so simple and powerful that it should be widely applied, and we refer to this principle as:
\begin{tcolorbox}[colback=green!5!white,colframe=green!75!black, enlarge top by=1cm, enlarge bottom by=1cm]
\textbf{BCI-AntiLink}: In the different data streams (brain signals, behavioral data, stimuli, multimodal signals), any plaintext mutual identifiers should be removed before storage. This includes subject identifiers or common time stamps, which would allow linking corresponding files.  
\end{tcolorbox}

\section{Separate}

Another important privacy-by-design strategy is to logically or physically Separate the processing of personal data. This strategy has two associated tactics, \emph{Isolate} and \emph{Distribute}. \emph{Isolate} refers to logical and physical separation even though databases or applications are still run on centralized hardware. In contrast, \emph{Distribute} refers to distributed processing over separate physical locations. As laid out by Hoepman, this latter tactic involves relying on the local equipment (in our scenario the local \acrshort{bci} hardware) of the data subject as much as possible, and centralized components (in our scenario the remote \acrshort{bci} components) as little as possible. Therefore, the Separate strategy is closely linked to \acrshort{ppml} techniques including \acrshort{smc} and federated learning which allow moving from centralized to decentralized solutions as discussed above (cf. Fig.~\ref{fig:mapping-hoepman} where we indicate this transition by the yellow arrow). 

\section{Inform}

The aim of the Inform Strategy is to ensure transparency about personal data processing. This includes transparency about what data are processed how and to what purpose, and the three tactics \emph{Supply}, \emph{Explain}, and \emph{Notify}. In our view, at the present developmental stage of \acrshort{bci} technology, this strategy is arguably the most important component of a \acrshort{bci} privacy strategy at all. On the level of the individual, transparency is a necessary precondition for exerting the right of informational self-determination and for users to make informed decisions. On the societal level, transparency is a precondition for verifying the legal and regulatory compliance of organizations (as put by Hoepman: \emph{\say{Sunlight is said to be the best of disinfectants.}}).

Transparency in the context of brain data and \acrshort{bci}s is especially challenging, as we do not have a natural intuition about the information contents of brain signal recordings and the related privacy issues, in contrast to, e.g., e-mails or photographs, where we have a much better understanding about the private information they contain about us. Therefore we argue that measures to implement an elaborate Inform Strategy is to ensure \verb|"BCI transparency"| and should be woven into the fabric of the whole \acrshort{bci} setup - as indicated by the blue box (Fig.~\ref{fig:mapping-hoepman}) encompassing all of the \acrshort{bci} components.

We propose \textbf{transparency as a key element of a privacy-preserving strategy for \acrshort{bci}s}, addressing (at least) the following two different levels of transparency:  
\begin{itemize}
    \item What is decoded in an online \acrshort{bci}? 
    \item Which information can be derived from the stored data? 
\end{itemize}

\subsection{What is decoded in an online BCI?}
To address the first level (\emph{What is decoded in an online \acrshort{bci}?}) we propose a transparency- oriented mode of operation for online \acrshort{bci}s. This will be a useful mechanism also to ensure that a \acrshort{bci} is only used for the stated and legitimized purpose. 

\bigskip
\bigskip
\bigskip

\begin{figure}[!b]
\begin{centering}
    {\includegraphics[scale=0.4]{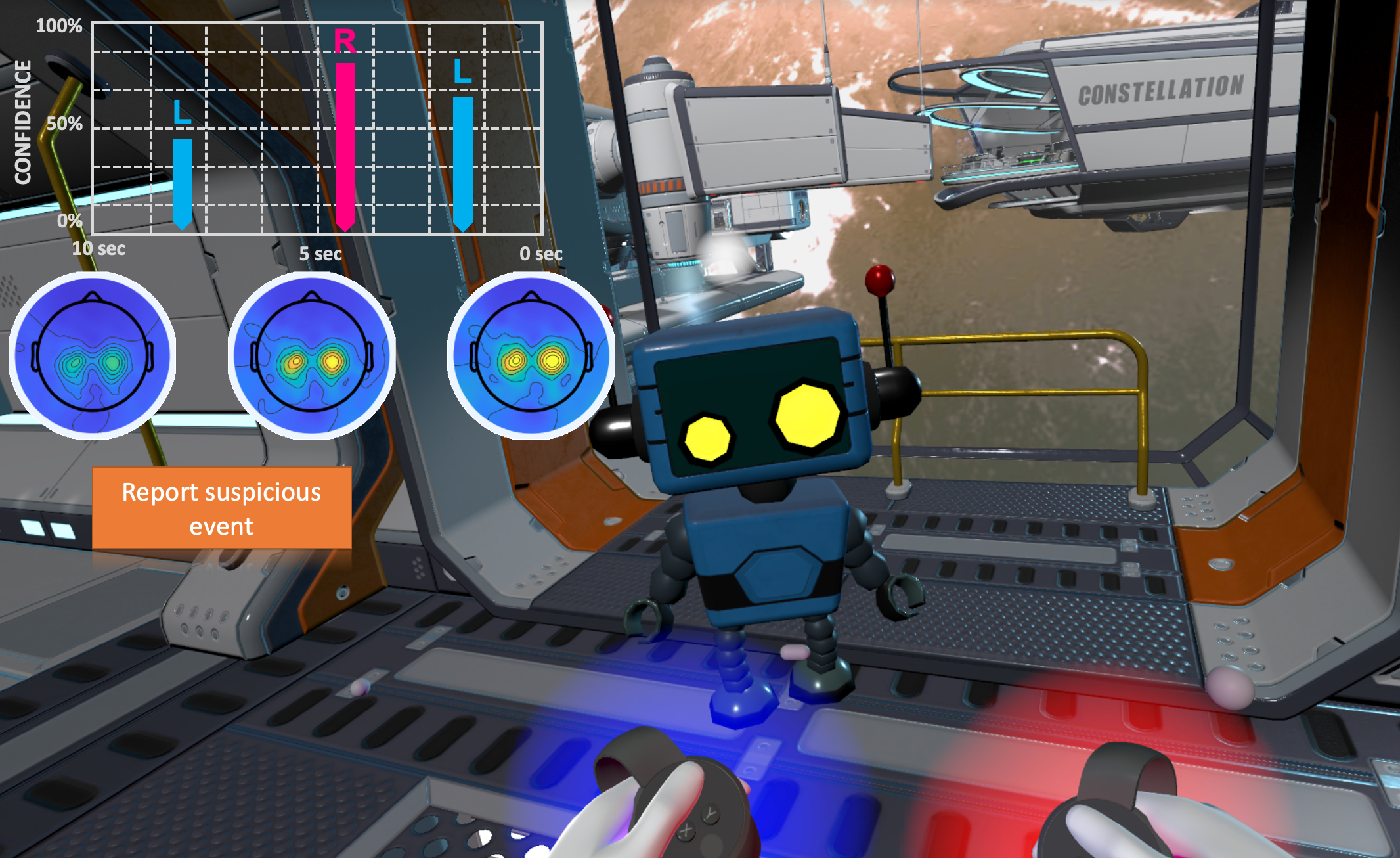}}
    \caption[Design sketch for BCI TransparentMode.]{\textbf{Design sketch for BCI TransparentMode} based on a BCI-controlled VR neurogame (currently developed by the authors and coworkers), where subjects use imagined right (R) and left (L) hand movements to interact with virtual robots. On the top left, the TransparentMode display components show decoded classes over time, bars scaled by model confidence of classification. Maps below would visualize an EEG map of the features important for the classification decisions. Subjects are hence able to verify that the decoded information makes sense given their behavior in the game and to report suspicious events. }%
    \label{fig:transparent-mode}
\end{centering}
\end{figure}

\begin{tcolorbox}[colback=green!5!white,colframe=green!75!black, enlarge top by=1cm, enlarge bottom by=1cm]
\textbf{TransparentMode for online BCIs:}\footnote{Note that the Inform Strategy addresses Unawareness Threats falling into the soft privacy category (Tab.1). Thus, transparency-oriented measures as described above are only helpful if the BCI provider (who would also be responsible for implementing, e.g. the transparency mode as described above) is trusted.}\footnote{Such a transparent mode might distract users in critical applications, therefore as stated, it would be important that it can always be switched off.}
\begin{itemize}
    \item Most \acrshort{bci}s today are based on decoders which are trained in a supervised manner, e.g., to infer right vs. left imagined hand movement from \acrshort{eeg}.
    \item For this supervised training, in most of the cases there is a human-interpretable, semantically meaningful class or regressor label (such as which hand was moved, which letter was selected, the level of tiredness). 
    \item \acrshort{bci}s should have a \textbf{transparent mode}, which can be arbitrarily switched on and off by the user, in which all decoded labels are displayed to the user online.
    \item The key advantage of such a transparent mode is that it allows the user to make sanity checks online. In the hand-movement decoding case, for example, the user can match the decoded labels to the imagined hand movement. For example, if the decoded labels do not match at all to the associated behavior, either the decoder is not working properly, or the decoder is in fact inferring something different than pretended. Thus, a transparent mode could not only contribute to uncover attempts to steal information online, or to uncover \verb|"fake BCIs"| which otherwise may hide behind a magic black box marketing.
    \item In addition, all decoded labels should by default be stored in a plaintext log file which is also accessible for the user.
\end{itemize}
\end{tcolorbox}

\begin{tcolorbox}[colback=magenta!5!white,colframe=magenta!75!black]
How can \textbf{interpretable/explainable \acrshort{ml}} be used to provide \textbf{online feedback}  helping to understand the decisions of the \acrshort{ml} model including uncertainties / model confidence, visualizing which signal components are used, etc., not only by domain experts, but also \acrshort{bci} users? 
\end{tcolorbox}

\subsection{Which information can be derived from the stored data? }
In the previous paragraph we addressed the issue of transparency of the online processing of data during \acrshort{bci} usage. Equally, or maybe even more important is transparency with respect to the offline processing of data collected during \acrshort{bci} usage. 

As a \acrshort{bci}-specific step towards more transparency with respect to offline processing of \acrshort{bci}-related brain data, we propose \textbf{BCI quality benchmarking (BCIQ-bench)}, a standardized benchmarking for addressing the fundamental issue that \acrshort{bci} users so far have no means to gain even a most basic insight into the informativeness of the brain signals recorded and stored in the context of the \acrshort{bci} usage. It has been stated that \emph{\say{(EEG), even with low number of sensors, is an extremely rich signal}}~\cite{stopczynski_privacy_2014} - but can we a priori be sure of this? And what does \emph{extremely rich} mean exactly? BCIQ-bench (or other similar benchmarking approaches) could be implemented \verb|"here and now"| to address these questions without the need for complex methods development, and yet could substantially help promote informed user decisions. 

\begin{tcolorbox}[colback=green!5!white,colframe=green!75!black, enlarge top by=1cm, enlarge bottom by=1cm]
\textbf{BCI.Q-Bench}: Age and gender are two widely available labels which can be decoded from brain data, in particular there are good baselines about the performance one can expect from EEG data~\cite{engemann_reusable_2021, siddhad_efficacy_2022}, and there are efforts to promote standardized and reusable benchmarks for example for age decoding from \acrshort{eeg}~\cite{engemann_reusable_2021}. Requiring \acrshort{bci} providers to publish age and gender decoding accuracies for the data recorded with their \acrshort{bci} systems would provide users with at least basic information. \acrshort{bci} benchmarking, prospectively, could be extended to other standardizable, privacy-related tasks, to provide users with a solid overview about how informative (or not) the recorded brain signals are.   
\end{tcolorbox}

\begin{tcolorbox}[colback=magenta!5!white,colframe=magenta!75!black]
How can the \textbf{amount of personal information decodable from brain signals} with a given set of \acrshort{ml} methods be reliably assessed? Public benchmarks as above are an important first step, but over time such benchmarks tend to be overfitted and if all labels are public, are open to cheating. In the best case, they reveal only the “tip of the iceberg”, as many aspects of personal information will not be covered by such benchmarks. Also, performance on offline benchmarks cannot substitute for evaluation in online applications. Competition (similar to the Cybathlon\footnote{https://cybathlon.ethz.ch/de}) could offer a way ahead. 
\end{tcolorbox}

\section{Control}
The Control Strategy fulfills the fundamental role to ensure that users have control over the processing of their personal data. The control tactics are \emph{Consent}, \emph{Choose}, \emph{Update}, and \emph{Retract}. Logically the Control strategy builds on the Inform aspects as discussed above, as it impedes the meaningful control of a system if one has no understanding of it. Therefore we argue that a major focus at the current development level of \acrshort{bci} technology should aim at increased transparency and promote better understanding, including \acrshort{bci} competence of the users, explainability of the used \acrshort{ml} methods, etc., and that appropriate control strategies can be built on the basis of such transparency and improved understanding. 

This being said, there also is a range of \textbf{basic control functions} which can already be implemented in the \verb|"here and now"| of the current stage of development, without needing to wait for the development of a better \acrshort{bci} transparency fundament. For example, to ensure user control over the data, \acrshort{bci}  providers should have strong data governance (see~\ref{sec:regulatory}) and provide detailed information to users about how and where the data are stored, and delete the data upon request. Informed consent should be systematically obtained from the \acrshort{bci} users, with possibility of withdrawal.

\subsection{No “Always-On BCIs”}
One very simple yet apparently not entirely self-evident example of an effective \acrshort{bci} control is that \acrshort{bci}s should have an \textbf{On/Off switch}, as called for by~\cite{ibm-paper, internetofthinks}. This switch may, if necessary, be incorporated both in software and hardware \acrshort{bci} design, to ensure fine-grained temporal control of the device by the user. This is important, as there are known privacy and security threats with \verb|"always on"| devices~\cite{always-on}; studies highlighted the feasibility of decoding private information from long-term \acrshort{bci} usage (all -day recordings), especially in scenarios when the main activity for which \acrshort{bci} is intended (e.g. gaming) is briefly stopped and a user is, for example, using an online banking system while still wearing the continuously recording \acrshort{bci}~\cite{neupane_2017, neupane_brain_2019}. Privacy risks may also arise when combining brain data from an always-on \acrshort{bci} with other types of user activity, such as keystroke dynamics~\cite{neupane_brain_2019}. Also, similar to webcams, \acrshort{bci}s should clearly indicate to the user whether they are in recording mode or not.

\begin{tcolorbox}[colback=green!5!white,colframe=green!75!black, enlarge top by=1cm, enlarge bottom by=1cm]
\textbf{Not-AlwaysOn-BCI}: On principle, unless there are medical or safety reasons which would require continuous operation, \acrshort{bci}s should not be designed as always-on devices, have an On/Off switch, and clear indicators whether they are in recording mode or not.  
\end{tcolorbox}

\subsection{Ensuring user autonomy}\label{sec:userautonomy}

Beyond brain privacy, it is also crucial to ensure that users maintain control over actions mediated via active \acrshort{bci} devices. We envisage this functionality as a final \verb|"safety net"| to catch as early as possible all \acrshort{bci} actions which are against the user’s intent before they unfold (\acrshort{bci} errors due to misclassification or malicious activities). Equally, user autonomy is related to the notion of liability, therefore we also aim here to prevent \verb|"cheating"| and ensure that \acrshort{bci} outcomes are genuine and reflect true user intent. Thus the \acrshort{bci} device cannot be blamed if the adversary is the user him/herself (non-repudiation). Note that in the case of closed-loop \acrshort{bci} /stimulation the notion of user autonomy gets another level of complexity and requires separate considerations.

Furthermore, we propose two classes of  functions for ensuring user autonomy: The first class comprises explicit \emph{alert} and \emph{emergency functions} or intervention options, e.g., the \acrshort{bci} analog to an \textbf{emergency stop function}. For example, in one of our previous online \acrshort{bci}s for robot control~\cite{kuhner_service_2019}, users could stop the \acrshort{bci}-controlled robot actions any time by an imagined command. Alert functions will notify users in case of detection of suspicious activities within the \acrshort{bci} system. Emergency functions are intended to also notify trusted parties (for example, caregivers) about device malfunctioning and halt the device if possible. The mandatory On/Off switch as mentioned above would also give users more granular temporal control of the device.

The second class of functions to detect \acrshort{bci} behavior violating the user’s intent is based on \emph{brain responses caused by unintended actions}. Decoding such implicit error-related brain responses, like the \acrfull{ern}, online provides an elegant approach to ensuring that users maintain control over the \acrshort{bci}~\cite{volker_dynamics_2018, volker_intracranial_2018}. Notably, explicit and implicit methods could also be combined within one and the same \acrshort{bci} application. Further, data-driven anomaly detection may help to identify outliers in the \acrshort{bci}-controlled behavior and ask for additional confirmation (\emph{\say{Do you really want olives on your pizza, you never eat them otherwise?}}). Thus for specific \acrshort{bci} use cases, it will be important to consider such approaches and how they can be optimally engaged/combined to ensure user autonomy over their \acrshort{bci}. 

\begin{tcolorbox}[colback=green!5!white,colframe=green!75!black, enlarge top by=1cm, enlarge bottom by=1cm]
\textbf{BCI-AutonomyGuard 1.0}: Including an emergency stop function to ensure the possibility of a user veto over their active \acrshort{bci}.
\end{tcolorbox}

\begin{tcolorbox}[colback=magenta!5!white,colframe=magenta!75!black]
\textbf{BCI-AutonomyGuard 2.0}: How can we use an optimized combination of explicit (emergency stop function), intrinsic (error-related brain responses), and data-driven (anomaly detection) methods to ensure user autonomy over their active \acrshort{bci} at all times?
\end{tcolorbox}

\section{Enforce and Demonstrate}
Finally, the Enforce and Demonstrate strategies address how to anchor privacy policies as part of the organizational culture, how they can be propagated by management, as well as how to demonstrate compliance to privacy regulations towards the data protection authorities. These strategies are thus less concerned with the design of secure and privacy-preserving \acrshort{bci}s, which is the main topic of our framework, but these strategies clearly need to be embedded on on the legal and regulatory level. Thus, in the following, we will review some legal and regulatory aspects related to \acrshort{bci} privacy and security considerations.  
\subsection{Regulatory and legal aspects of BCIs}\label{sec:regulatory}
Since the Charter of Fundamental Rights of the European Union became binding in 2009, data protection has acquired the status of a fundamental right (Article 8) throughout the EU. The implementation of the \acrshort{gdpr} (\acrfull{dsgvo} in Germany) has subsequently positioned the EU as a global leader in data protection regulation. Furthermore, data protection is of particular relevance in Germany - not only against the rapid development of information technology, but also because of historical experiences with political regimes that collect information to suppress citizens. For a recent analysis of this German perspective, see~\cite{iphofen_privacy_2021}. Another study reported Germans stood out as the only nationality placing more value on the privacy of health data than credit card information~\cite{harvard-business}. 

In Germany, a number of different legal texts address the security of patient data for medical facilities or data protection in general. In addition to the \acrshort{dsgvo} and the \acrfull{bdsg}, there are also regional state data protection laws (e.g., the Datenschutzgesetz Nordrhein-Westfalen). Application of general principles to brain data may require special considerations, such as the Principle of data avoidance and data economy of the \acrshort{bdsg}. Data anonymization and limitation is a special challenge in the context of \acrshort{bci} devices, as discussed above. The detailed legal situation with respect to \acrshort{bci} technology in Germany has been recently reviewed by Martini and Kemper~\cite{martini_cybersicherheit_2022}. The authors conclude that on the one hand, our current legal system already formulates individual requirements for the cybersecurity of \acrshort{bci}s, but that it lacks a complete regulatory concept that adequately guarantees the security of the applications. Medical device law in Germany regulates security with respect to patient safety, the situation for non-medical \acrshort{bci}s is however substantially different and is characterized by various open regulatory gaps. Martini and Kemper argue that the new EU directives on digital products may at least close these gaps to some extent, as companies are obliged to provide software updates in order to maintain contractual compliance of digital products, which would at least encompass the \acrshort{bci} software components. 

The EU Commission's draft regulation on \acrfull{ai} also provides for strict requirements for safety of \acrshort{ai} systems. On the European level, in Article 52 of the EU \acrshort{ai} act, the proposed EU \acrshort{ai} law regulates \emph{\say{Transparency obligations for certain AI systems}}. This article states that users of \emph{\say{an emotion recognition system or a biometric categorisation system shall inform of the operation of the system the natural persons exposed thereto.}} This article could thus be applied at least to some \acrshort{bci} systems using \acrshort{ai} technology in the context of emotion recognition. Given the pivotal role of transparency for the development of trustworthy \acrshort{bci}s, we consider that a discussion on \textbf{extended transparency obligations} for \acrshort{ai}-\acrshort{bci} systems also beyond the narrow topics of emotion recognition and biometric categorization could be fruitful.

\subsection{Do we need a special governance framework for BCI data? }
On a general governance level, we have recently proposed a \textbf{Governance Framework for Brain Data}~\cite{ienca_towards_2022}. In this framework, we have identified distinctive ethical and legal implications of brain data acquisition and processing, and have outlined a multi-level governance framework for brain data. The rationale of this framework is aimed at maximizing the benefits of brain data collection and further processing for science and medicine whilst minimizing risks and preventing harmful use. The framework consists of four levels of regulatory intervention: binding regulation, ethics and soft law, responsible innovation, and human rights. This framework considers brain data in general, including brain data obtained through \acrshort{bci}s. Due to its comprehensive nature, we are confident that this Governance Framework for Brain Data also adequately covers the \acrshort{bci} data case, and that on this generic level, at least for the time being, no additional \acrshort{bci}-specific considerations and specifications are obviously necessary.

%% file: chapters/7-application.tex
\chapter{Application to Specific BCI Use Cases and Contexts}\label{chap:application}
Up to here, we have described a general Framework for Brain-Privacy-Preserving Cybersecurity in \acrshort{bci} Applications. How can this framework be applied to specific use cases?  Such use cases may include a wide range of application scenarios, and imply different security and regulatory requirements, broad \acrshort{bci} categories include: 
\begin{itemize}
    \item Clinical vs. non-clinical,
    \item Invasive vs. non-invasive, 
    \item Commercial \acrshort{bci} product development vs. academic research and lab prototypes. 
\end{itemize}
Importantly, different contexts will require a \textbf{different depth of analysis and considerations}. Commercial \acrshort{bci} products may require a fine-grained, full-fledged analysis. For academic research and lab prototypes, our work may offer a more high-level, systematic, conceptual framework even in cases where a detailed, step-by-step analysis is not feasible or useful. 

Basically, application of our framework to specific use cases will proceed in the same steps as described above, namely threat modeling, risk assessment, and privacy engineering. Starting with a \acrshort{dfd}, it is important to emphasize that not all \acrshort{bci} use cases will necessarily have all the components we have mapped out in our extended \acrshort{bci} cycle and the corresponding \acrshort{dfd}. For example, a research-grade implantable \acrshort{bci} system for communication in a locked-in patient may not have remote components; \acrshort{bci} model training in this case may proceed purely within-subject but longitudinally over a long time period. On the other hand, it is also imaginable that for example a neurogaming \acrshort{bci} might run all functionality except for the core \acrshort{bci} cycle in the cloud, thus omitting the extended core according to our terminology. Future \acrshort{bci}s may also include additional components not envisaged in our schematic, or more fine-grained \acrshort{dfd}s may be called for. 

In the next step, threat modeling and risk assessment is performed. In contrast to our general framework, however, not only general risk potential but also specific risks for a concrete use case can be assessed, because the specifics of the \acrshort{bci} system, such as the information content of the recorded brain data, are known (or else should be determined). For example, a risk assessment of a low-cost \acrshort{bci} with a few channels of non-invasively recorded signals will result in much lower risk assessments compared to a high-end system potentially providing hundreds or even thousands of much more informative recording channels. 

Such differences in the estimated risk which have profound consequences for the threat mitigation will be implemented in the following step, moving through the 8 privacy strategies as described in detail above. For example, in our general architecture (Fig.~\ref{fig:mapping-hoepman}), we have assigned the \emph{Minimize}, \emph{Abstract} and \emph{Hide} strategies to all three processing nodes of our general \acrshort{bci} data flow schematic. However, in a low-privacy-risk \acrshort{bci} application, it might be deemed sufficient to implement the Minimize strategy for the \acrshort{bci} core, and the Hide strategy on the remote level. Implementation of sophisticated \acrshort{ppml} techniques may be waived considering an overall low risk. In contrast, in a high-risk and safety-critical \acrshort{bci} application, the full range of technically possible defenses across all \acrshort{bci} components may be required.

Finally it is important to emphasize that we have made specific choices concerning the tools we selected for each of these steps; we have discussed the motivation of these choices above, such as for example the aim of achieving \acrshort{gdpr} compliance. However, the development of systems engineering methodologies is a dynamic field, and more suitable tools may exist in the foreseeable future, or already now in other legal contexts, e.g. the US data privacy framework \acrfull{hipaa}. Thus the choice of specific tools for threat modeling, risk assessment, and privacy engineering should be reviewed and if necessary they should be adapted to the specific legal and regulatory context. For example the LINDDUN framework, and similarly also STRIDE, is designed to be independent of the specific risk assessment technique that is used, providing the analyst freedom of choice, if needed, thus contributing to the flexibility and extensibility of the systematic approach to \acrshort{bci} privacy and security as advocated here.

%% file: chapters/8-conclusions.tex
\chapter{Conclusions and Outlook}\label{chap:conclusions}
\section{Conclusions}
Here, we have presented a framework for preserving privacy and cybersecurity in \acrshort{bci} applications. We started out with a working definition of brain privacy based on concepts of informational privacy and self-determination. Then, we delineated the functional, hardware, and software scope of current and prospective \acrshort{bci} applications. We introduce a novel, extended model of \acrshort{bci} functionality, consisting of the core \acrshort{bci} cycle, an extended core, and global functionality. We discuss the different ways in which this functionality can be implemented in concrete \acrshort{bci} applications with embedded, local computing, and cloud components. Based on these considerations we derive a \acrshort{bci} \acrshort{dfd} as the key abstraction underlying the following steps of analysis.
\bigskip
\begin{figure}[h!]
\begin{centering}
    {\includegraphics[scale=0.3]{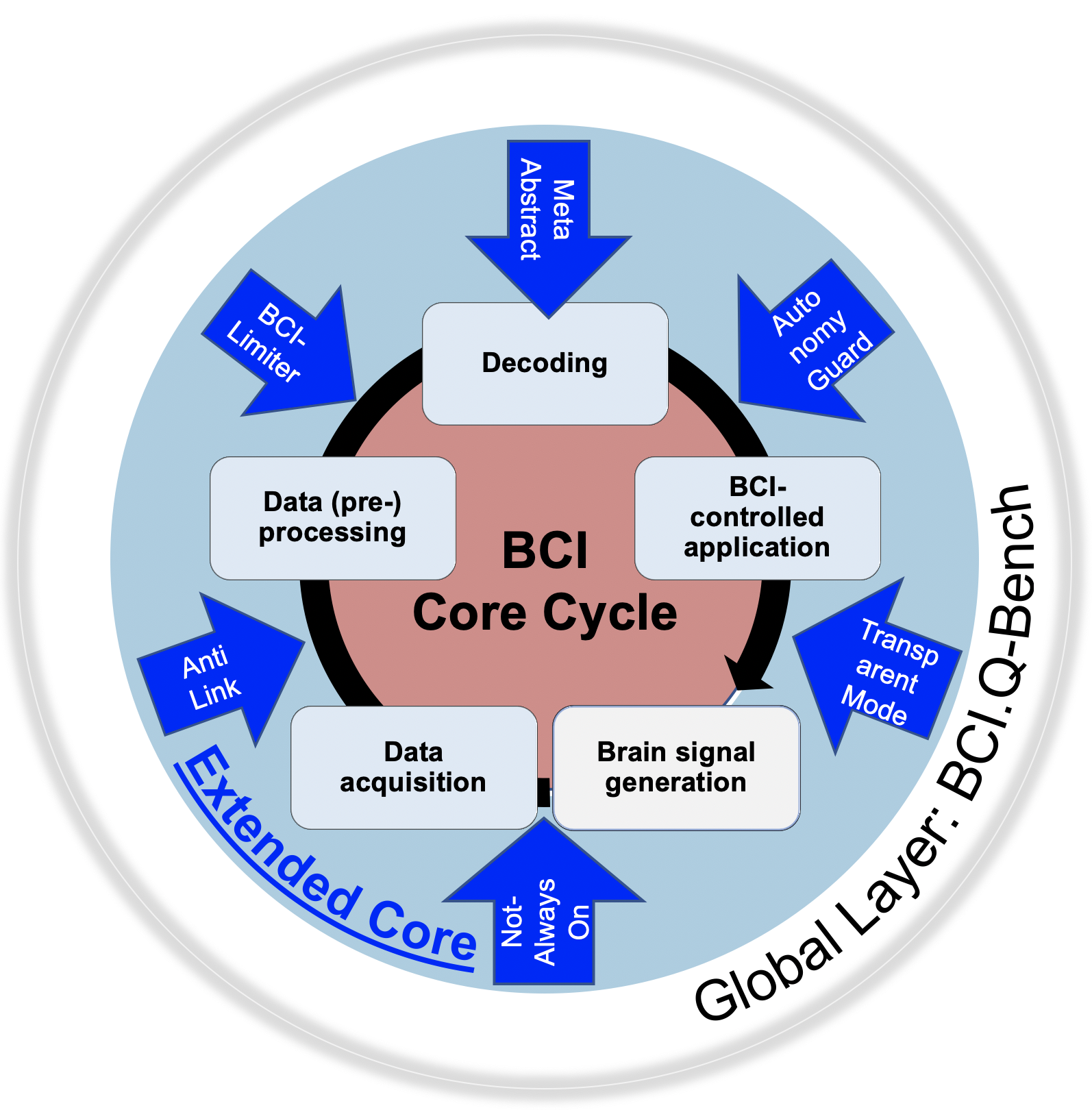}}
    \caption[Privacy-preserving \acrshort{bci} design features overview.]{\textbf{Privacy-preserving \acrshort{bci} design features overview.} Most of the features proposed within our framework could be implemented as part of the extended \acrshort{bci} core. We envision that such an extended \acrshort{bci} core might serve as a “\acrshort{bci} privacy shield” integrating multiple privacy-preserving mechanisms}%
    \label{fig:final-core}
\end{centering}
\end{figure}
 
We performed a systematic privacy threat modeling and risk assessment using the LINDDUN and \acrshort{owasp} methodologies, respectively, and systematically evaluated design patterns addressing these risks along the Hoepman  privacy design strategies. Thus, we delineate both a landscape of \acrshort{bci}-related privacy threats and risks as well as of mitigation strategies and tactics. Because our framework is entirely based on systematic methodologies, it can be also adapted to concrete \acrshort{bci} use cases. 

Our results here provide both a blueprint for brain privacy-preserving \acrshort{bci} architectures, and concrete, actionable design features. In Fig~\ref{fig:final-core}, we have mapped these features to our functional \acrshort{bci} overview. Most of the proposed features are naturally fitting to an implementation within the extended \acrshort{bci} core, namely the \acrshort{bci}-Limiter, \acrshort{bci}-AntiLink, TransparentMode, Not-Always-On-\acrshort{bci}, \acrshort{bci}-AutonomyGuard, and \acrshort{bci}-Meta-Abstract (\acrshort{bci}.Q-Bench as a benchmarking tool for data from multiple users also involves global functionality) -- highlighting the usefulness of distinguishing the extended core as a separate important functional \acrshort{bci} layer. We envision that such an extended \acrshort{bci} core might serve as a \verb|"BCI privacy shield"| integrating multiple privacy-preserving mechanisms.

\section{Outlook: Privacy and security across the BCI life cycle}
Ensuring privacy and security across the \acrshort{bci} life cycle brings additional challenges. Brain signals display nonstationarities on timescales from milliseconds to years. Dealing with these changes in the measurement streams of neuronal activity is a fundamental challenge in brain signal analytics and a topic of ongoing research. Thus consistent safety of \acrshort{bci} applications must be ensured under the presence of such nonstationarities in the brain signals. Specifically, robust and adaptive decoders should be designed to deal with this challenge. For example, in our previous work we implemented and successfully tested a first online \acrshort{bci} using adaptive \acrshort{dl}~\cite{kuhner_service_2019}, finding that the adaptivity of the networks was critical to high \acrshort{bci} performance. Adaptive algorithms, however, come with their own risks and challenges, for example, an adaptive decoder may become unstable or adapt to a suboptimal \acrshort{eeg} feature. On the other hand, neuronal variability may impede, e.g., side-channel attacks on \acrshort{bci} systems~\cite{lange_side-channel_2017}; thus variability may even help to make \acrshort{bci}s safer. Extending our current framework to encompass the entire \acrshort{bci} life cycle would therefore be a perspective for future work.